\documentclass[12pt]{iopart}

\usepackage{qcircuit}
\usepackage{graphicx}
\usepackage{subcaption}

\begin{document}

\title{Robust Two-Qubit Gates Using Pulsed Dynamical Decoupling}

\author{Patrick Barthel$^1$, Patrick H. Huber$^1$, Jorge Casanova$^{2,3}$, I\~nigo Arrazola$^{2,4}$, Dorna Niroomand$^1$, Theeraphot Sriarunothai$^1$, Martin B. Plenio$^5$, Christof Wunderlich$^1$ }
%\email[]{barthel@physik.uni-siegen.de}
%\homepage[]{Your web page}
%\thanks{}
%\altaddress{}
\address{$^1$Department of Physics, School of Science and Technology, University of Siegen, 57068 Siegen, Germany}

\address{$^2$Department of Physical Chemistry, University of the Basque Country UPV/EHU, Apartado 644, 48080 Bilbao, Spain}
\address{$^3$EHU Quantum Center, University of the Basque Country UPV/EHU, Leioa, Spain}

\address{$^4$Vienna Center for Quantum Science and Technology, Atominstitut, TU Wien, 1040 Vienna, Austria}
\address{$^5$Institute of Theoretical Physics and IQST, Albert-Einstein-Allee 11, Ulm University, 89069 Ulm, Germany}
%

%\ead{submissions@iop.org}
\vspace{10pt}
\begin{indented}
\item[]\date{\today}
\end{indented}

\begin{abstract}
We present the experimental implementation of a two-qubit phase gate, using a radio frequency~(RF) controlled  trapped-ion quantum processor. The RF-driven gate is generated by a pulsed dynamical decoupling sequence applied to the ions' carrier transitions only. It allows for a tunable phase shift with high-fidelity results, in particular a fringe contrast up to $99_{-2}^{+1}\%$ is observed in Ramsey-type measurements. We also prepare a Bell state using this laser-free gate.
The phase gate is robust against common sources of error. We investigate the effect of the excitation of the center-of-mass~(COM) mode, errors in the axial trap frequency, pulse area errors and errors in sequence timing. The contrast of the phase gate is not significantly reduced up to a COM mode excitation $<20$~phonons, trap frequency errors of +10\%, and pulse area errors of -8\%.
The phase shift is not significantly affected up to $<$ 10~phonons and pulse area errors of -2\%. 
Both, contrast and phase shift are robust to timing errors up to -30\% and +15\%.
The gate implementation is resource efficient, since only a single driving field is required per ion. Furthermore, it holds the potential for fast gate speeds (gate times on the order of $100~\mu$s) by using two axial motional modes of a two-ion crystal through  improved setups.
%~\cite{Arrazola2018}.
\end{abstract}


\section{Introduction\label{sec:Introduction}}
Trapped atomic ions are one of the leading platforms for realizing quantum computers~(QC). High-fidelity two-qubit entangling gates and Bell state preparation experiments have been realized in these systems using laser light to coherently drive qubit resonances and their motional sidebands ~\cite{Blatt2008, Benhelm2008, Ballance2016, Gaebler2016, Kaufmann2017, Landsman2019, Clark2021}. RF-controlled trapped ions promise quantum processors that are less challenging to scale up, since they do not rely on complex and demanding laser systems but use highly developed and easily accessible RF technology instead~\cite{Wunderlich2001, Ospelkaus2008,Johanning2009,Timoney2011, Ospelkaus2011, Khromova2012}. In RF-based trapped-ion QC, single-qubit rotations have been carried out  with high fidelity~\cite{Brown2011, Harty2014} and unrivaled low crosstalk~\cite{Piltz2014}, while two-qubit gate fidelities are on par with laser-based approaches. Recently, various approaches for robust two-qubit gates, as well as maximally entangled states with fidelities of 0.9977 have been reported in RF-based trapped ions~\cite{Harty2016, Webb2018, Zarantonello2019, Srinivas2021}, mostly relying on the application of several driving fields on or near the motional sidebands of each ion.

A versatile technique to protect quantum gates against decoherence is dynamical decoupling~(DD), based on the application of additional continuous driving fields, sequences of RF pulses or a combination of both~\cite{Timoney2011, Bermudez2012, Piltz2013, Puebla2016, Puebla2017, Manovitz2017, Webb2018, Arrazola2020}, generalizing earlier concepts for laser-driven trapped ions~\cite{Jonathan2000}. In this work, the gate is generated by the DD sequence by tuning its pulse timings to match the ion motion. Pulses are applied only to the carrier transitions of a two-ion crystal~\cite{Arrazola2018}.
By tuning pulse timings to the period of the axial motional modes of the crystal, a tunable conditional phase shift is generated between the ions, while rendering the evolution robust to several major sources of gate errors.

The paper is structured as follows. First, the experimental setup is outlined, before the gate mechanism and its derivation are described.
Then, measurement results are presented for the conditional phase shift and the  preparation of a Bell state. Furthermore, the robustness of the gate to motional excitation, errors in axial trap frequency, pulse areas, and sequence timings are reported.

\section{Experimental setup\label{sec:Setup}}
\begin{figure}
\centering
	\includegraphics[width = 0.45\columnwidth]{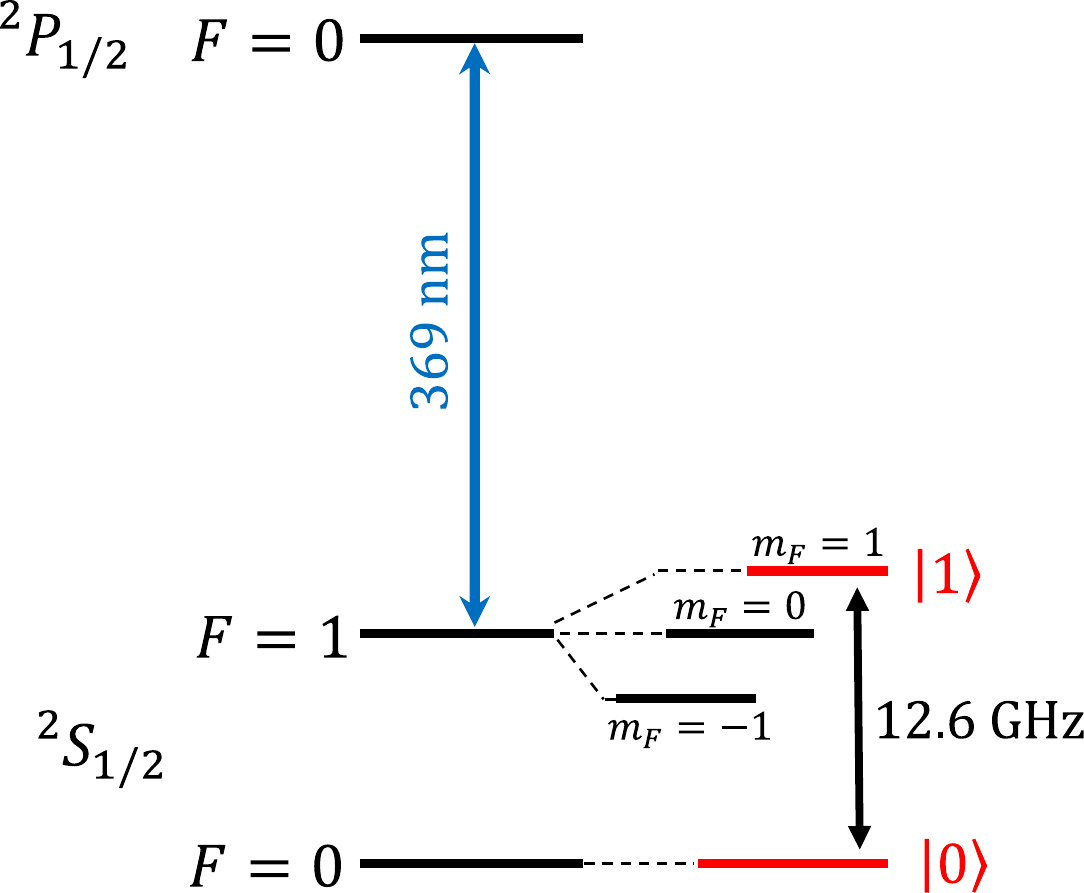}%
	\caption{\label{fig:Yb}Partial energy level structure of $^{171}\rm Yb^+$ including the hyperfine and Zeeman sublevels of the atomic ground state. Qubits are formed by the $|^2\rm S_{1/2}, \rm F = 0\rangle$ and the $|^2\rm S_{1/2}, \rm F = 1$, $\rm m_{\rm F} = 1\rangle$ levels (marked in red) with an energy splitting corresponding to $12.6$~GHz. Coherent operations are performed by RF radiation of the respective frequency. Incoherent driving for cooling, state preparation and read-out uses laser light at $369.5$~nm, driving the  $|^2\rm S_{1/2}, \rm F = 1\rangle \leftrightarrow |^2\rm P_{1/2}, \rm F = 0\rangle$ transition. Two additional optical transitions for repumping out of long-lived meta-stable states are not shown.}
\end{figure}

The two-qubit system used in this work is provided by a set of two $^{171}$Yb$^+$ ions trapped in a macroscopic linear Paul trap with radial and axial trapping frequencies of about $2\pi\times 380$~kHz and $2\pi\times 120$~kHz respectively~\cite{Piltz2016SciAdv}. The ions are Doppler cooled by laser light near 369.5~nm, which is also used for state read-out and preparation. Additional laser light near $935$~nm and $638$~nm provides repumping from long-lived states back into the cooling cycle~\cite{Bell1991, Gill1995}. Figure~\ref{fig:Yb} depicts a partial energy level structure of $^{171}$Yb$^+$. The ground state hyperfine levels used as qubit states are $|0\rangle  \equiv |^2\rm S_{1/2}, \rm F = 0, \rm m_{\rm F} = 0 \rangle$ and 
$|1\rangle \equiv |^2\rm S_{1/2}, \rm F = 1, \rm m_{\rm F} = 1\rangle$.

All coherent operations on the qubit states are carried out by applying global RF radiation near $12.6$~GHz to the ion crystal. Two permanent magnets held by the endcap electrodes generate a magnetic field gradient of $19$~T/m along the trap axis (z-axis), allowing for individual addressing of each ion in frequency space enabled by invidual Zeeman shifts. In addition, coupling of the ion's internal and motional states is provided via Magnetic Gradient Induced Coupling (MAGIC)~\cite{Wunderlich2001, Johanning2009, Khromova2012}. 

Here, single qubit operations are carried out with a Rabi frequency of $2\pi\times31$~kHz, i.e. $\pi$-pulse times of $16~\mu$s. The axial center-of-mass~(COM) of the common motion of the ion crystal at frequency $\nu_{1} = 2\pi\times120$~kHz is used for multi-qubit interaction.

Cooling beyond the Doppler limit is achieved by RF sideband cooling~\cite{Sriarunothai2018}. RF radiation drives a qubit's red sideband transition in addition to the laser at 369.5~nm, blue-detuned by 2.1~GHz using acousto-optical modulation. Ion motion within the harmonic trapping potential can be cooled down close to its ground state, with typically about 0.5~phonons.
State preparation is performed by optical pumping using the ${|^2\rm S_{1/2}, \rm F = 1\rangle \leftrightarrow |^2\rm P_{1/2}, \rm F = 1\rangle}$ transition, which yields a well-defined initial state $|0\rangle$ with a preparation efficiency ${\geq0.9975}$~\cite{Piltz2016}. 

Read-out of the qubit state is based on resonance fluorescence obtained by driving the $|^2\rm S_{1/2}, \rm F = 1 \rangle\leftrightarrow |^2\rm P_{1/2}, \rm F = 0\rangle$ transition with laser light near 369.5~nm. The qubit state is projected to the $\{|0\rangle$, $|1\rangle\}$ basis, which is distinguished by the amount of fluorescence photons gathered with an electron-multiplying charge-coupled device (EMCCD) camera. A double-threshold analysis is used to distinguish between the dark $|0\rangle$ and bright $|1\rangle$ states with fidelities of 99.5\% and 98.5\% respectively~\cite{Kaufmann2018}. These values summarize state preparation and measurement~(SPAM) errors, and all experimental results are corrected accordingly.

\section{Gate mechanism}
\label{sec:GateMechanism}
\begin{figure}[hb]
\centering
	\includegraphics[width = 0.4\columnwidth]%
	{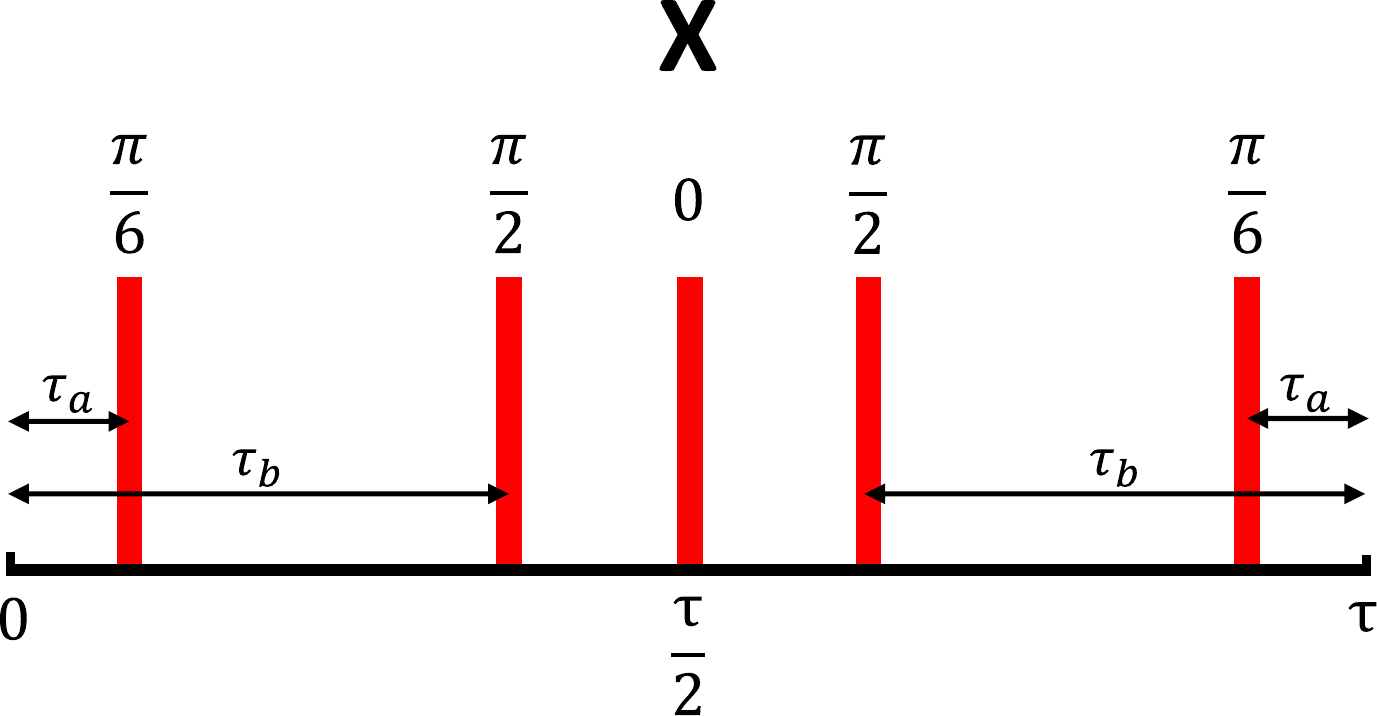}%
	\hspace{1cm}
	\includegraphics[width = 0.4\columnwidth]%
	{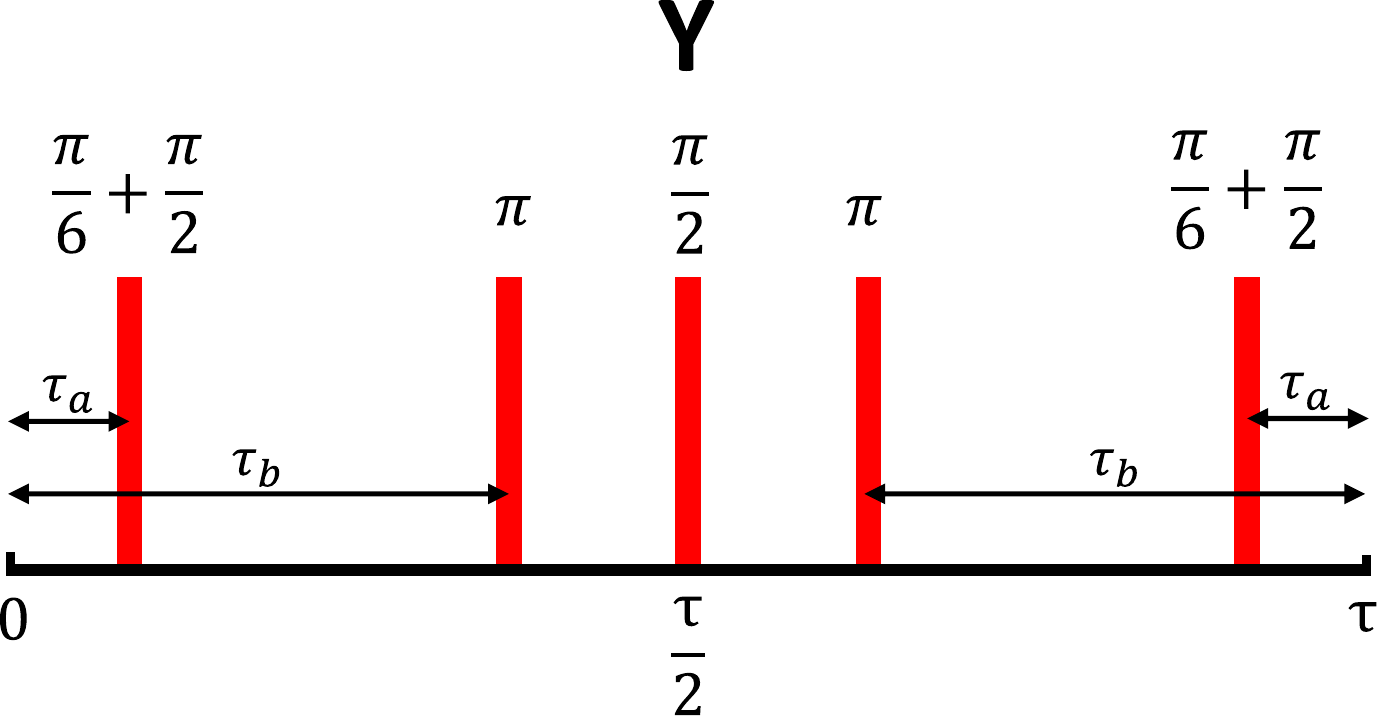}%
	\caption{\label{fig:AXY_Blocks}Building blocks of the adaptive XY decoupling sequence. Each block consists of a series of five non-equidistant $\pi$ pulses, distributed by the block duration $\tau$ and the pulse positions $\tau_a$ and $\tau_b$. Numbers above the pulses indicate the respective phases within each block, which provide robustness to pulse errors.}
\end{figure}

As detailed in Ref.~\cite{Arrazola2018}, the quantum gate is designed based on the adaptive XY~(AXY) sequence~\cite{Casanova2015} that was initially developed for quantum control of electronic spins in solids. A similar construction has been used for quantum sensing purposes in Nitrogen vacancy centers \cite{Hernandez2018, Unden2019}. The AXY sequence is a series of DD pulses applied to the carrier transitions of a two-ion crystal. Tunable pulse timings allow for adaption to the motional mode frequencies of the ions, providing the necessary coupling between qubit states and motional states.
Two blocks consisting of five~RF~$\pi$-pulses each, serve as building blocks of the DD sequence. As sketched in Fig.~\ref{fig:AXY_Blocks}, three tunable parameters $\tau_a, \tau_b$ and $\tau$ encode the time distance between $\pi$-pulses and determine block timings. Phases of $\varphi_x = \{\frac{\pi}{6}, \frac{\pi}{2}, 0, \frac{\pi}{2}, \frac{\pi}{6} \}$ and $\varphi_y = \varphi_x + \frac{\pi}{2}$ determine the rotation axes on the Bloch sphere and yield a self-correcting characteristic, i.e. robustness to pulse errors. Due to the $\frac{\pi}{2}$ phase difference, the two blocks are labeled X and Y.

If sequence timings are set such that the rates with which pulses are applied match the period of the motional modes of the ion crystal, a phase shift is generated between the ions. Namely the COM mode is selectively addressed via  the block length $\tau$, while the pulse positions, determined by $\tau_a$ and $\tau_b$ can be tuned to match with the breathing mode frequency. Using calculations and numerical simulations, the sequence is adjusted to decouple the electronic degrees of freedom from the motion at the gate time $t_{\rm gate}$, while a nonzero phase shift $\Phi$ is  generated~\cite{Arrazola2018}, resulting in the gate
\begin{equation}
	U = e^{i\Phi(t)\sigma_z\sigma_z}.
	\label{eq:PhaseGate}
\end{equation} 
In this notation a factor of $\frac{1}{2}$ is absorbed into the gate, i.e. the observed phase shift is 
\begin{equation}
	\Phi' = 2\Phi.
	\label{eq:PhaseShift}
\end{equation}
Numerous combinations of $\{\tau, \tau_a,\tau_b\}$ are possible, allowing for a tunable phase shift at the end of the gate. Here, we focus on the special case of a $\Phi = \frac{\pi}{4}$ gate, which can be used for CNOT and entangling operations as shown in the following.

The possibility to match the gate parameters to both modes of the ion motion simultaneously allows for higher gate speeds than single-mode gates~\cite{Manovitz2017}. Spin-motion coupling is generated by the magnetic field gradient, which hence determines the speed of the gate. Since RF driving is applied to the ions' carrier transitions only, large couplings to both vibrational modes~(COM and breathing mode) require large Rabi frequencies. 

Here, a DD sequence with a suitable number of pulses to protect the experimental system against decoherence is constructed by concatenating $m$ of X and Y blocks (see \cite{Arrazola2018} for details). In the experiments presented here, a total number of 80 - 160 pulses is used, which yields sequences of $m = 16$ or $m = 32$ blocks (AXY-16 and AXY-32 respectively). The corresponding gate times are given by $m\tau\approx 6$~ms.

Our current setup does not allow to fully exploit the speed-up that would result from the simultaneous use of multiple modes, since the Rabi frequency is much smaller than the mode frequencies, and consequently, these are not excited. We expect to fully exploit this feature in future setups with higher Rabi frequencies. The gate is very resource efficient, as it can be implemented with a single driving field per ion and even a single global field is sufficient, if pulses are applied sequentially to both ions. For many qubits, signal generators reach their capacity limits, in terms of amplitude resolution, storage capacity, and amplifier power. Thus it is beneficial to minimize the number of RF fields for scaling up to larger systems. Here, we implemented the gate with one RF driving field per qubit.

Following the derivation in Ref.~\cite{Arrazola2018}, the two-ion Hamiltonian
\begin{equation}
	H = \sum_{j=1}^{2}\frac{\omega_j}{2}\sigma^z_j  + 
	\sum_{k = 1}^{2} \nu_ka_ka^\dagger_k + 
	\sum_{j,k =1}^{2} \Delta_{jk}\left(a_k + a^\dagger_k\right)\sigma^z_j
\end{equation}
describes the system with qubit energies $\omega_j$, motional mode eigenfrequencies $\nu_k$ and the respective coupling constants $\Delta_{jk}$, using the Pauli matrix $\sigma_z$ and creation and annihilation operators $a$ and $a^\dagger$ (here, $\hbar$ is set to unity).
RF driving of both of the atomic carrier transitions with Rabi frequencies $\Omega_j$ and phases $\varphi_j$ is expressed as
\begin{equation}
	H_{C}(t)  = \sum_{j=1}^{2}\Omega_j(t)\left(\sigma^x_1 + \sigma^x_2\right)\rm cos(\omega_jt - \varphi_j),
\end{equation}
using the Pauli matrix $\sigma_x$.

In a suitable rotating frame with respect to qubit energies, motional frequencies and RF drivings, %See above Eq 3 
the interaction Hamiltonian reads
\begin{equation}
	H_{II}(t) = \sum_{j, k=1}^{2} f_j(t)\sigma^z_j\left(\Delta_{jk}a_ke^{-i\nu_kt} + \rm h.c.\right),
\end{equation}
taking the pulsed nature of the driving fields into account by the modulation functions $f_j = \pm 1$ for an even or odd number of pulses respectively.

Solving the Schrödinger equation for this Hamiltonian, a propagator $U$ is obtained with
\begin{eqnarray}
    \centering
	&U(t) &=  U_S(t)U_C(t),  \\	
	\textrm{where } &U_S(t) &= \rm exp\left\{\sum_{j, k=1}^{2}\left(
		\alpha_{jk}(t)a^\dagger_k - \rm h.c.\right)\sigma^z_j\right\} , \\
	\textrm{and } &U_C(t) &= \rm exp\left\{i\Phi(t)\sigma^z_1\sigma^z_2\right\}.
\end{eqnarray}
Here, the term $U_S$ contains the coupling between internal and motional states with the functions $\alpha_{jk}(t)=-i\Delta_{jk}\int_0^t dt'f_j(t')e^{i\nu_kt'}$, while the final term $U_C$ corresponds to the desired phase gate, with a phase shift $\Phi(t)$.

Simulations for experimental parameters but neglecting decoherence sources predict a gate with fidelities exceeding 99.9\% by tuning the AXY pulse sequence such that the block length $\tau$ is set with respect to the COM mode frequency $\nu_1$ and satisfies  $\nu_1\tau = 2\pi r$, with an integer $r$, ensuring $\alpha_{j1}(\tau) = 0$. Thus decoupling of the electronic degrees of freedom from the COM mode is achieved at integer multiples of $\tau$, while a conditional, nonzero phase shift $\Phi(\tau)\neq0$ remains.
Positions $\tau_a$ and $\tau_b$ of the pules within a block are then numerically optimized with respect to the breathing mode frequency $\nu_2$, to minimize $\alpha_{j2}(\tau)$, thus decoupling the breathing mode.
A set~$\left(\tau, \tau_a, \tau_b\right)$ can be chosen, such that  a desired phase shift $\Phi(\tau)$ is provided. Hence the phase shift is tunable by adjusting the sequence timings. 

\section{\label{sec:Results}Measurements and results}
We demonstrate the capabilities of the gate by measuring the resulting phase shift, generating a Bell state, and by investigating in detail the effect of common sources of error on the results. The measurement procedure is as follows. The ions are cooled close to the motional ground state, then initialized in the logical $|0\rangle$ state by optical pumping. This is followed by coherent manipulation using RF radiation, which implements the AXY DD sequence and additional single-qubit operations as given in the respective sections below.
Finally, laser illumination is used for state read-out.

\subsection{Phase shift}
Ramsey-type measurements are carried out to implement and verify the gate operation of the AXY sequence, by determination of the phase shift $\Phi'$ obtained according to Equation~\ref{eq:PhaseShift}.
\begin{figure}[hb]
\centering
	\includegraphics[width = 0.75\columnwidth]%
	{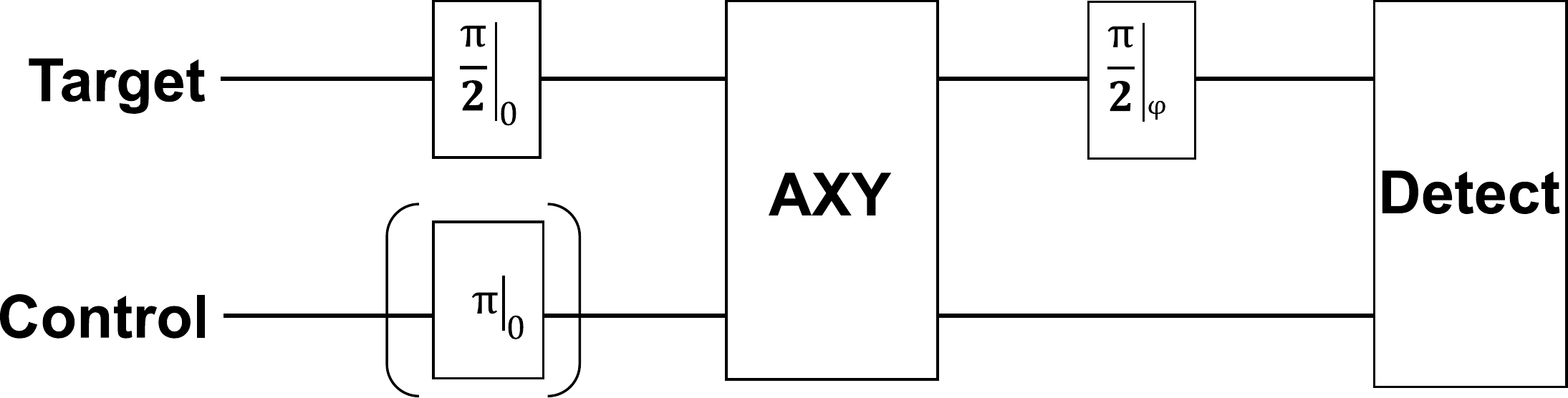}%
	\caption{\label{fig:AXY_Ramsey} RF sequence applied to a two-ion crystal for measuring the phase shift of the AXY sequence. A Ramsey measurement is performed on the target qubit, in the form of two $\frac{\pi}{2}$-pulses, while the control qubit is initialized to $|0\rangle$ or $|1\rangle$ by an optional $\pi$-pulse. In between the Ramsey pulses, the AXY gate sequence is applied to both qubits. State-selective detection by laser light reads out the final state of both qubits.}
\end{figure} 
Figure~\ref{fig:AXY_Ramsey} illustrates the RF sequence of two $\frac{\pi}{2}$-pulses to the target qubit with a fixed phase of $0$ for the first and a variable phase $\varphi$ for the second pulse. In between the pulses, the AXY decoupling sequence is applied to both qubits simultaneously with one RF field per ion. To measure the conditional response of the gate, the control qubit can alternatively be initialized to $|1\rangle$ by an additional RF $\pi$-pulse. After coherent RF manipulation, the laser at 369.5~nm is switched on to read-out the final state.

\begin{figure}[htb]
    \centering
	\includegraphics[width = 0.6\linewidth]{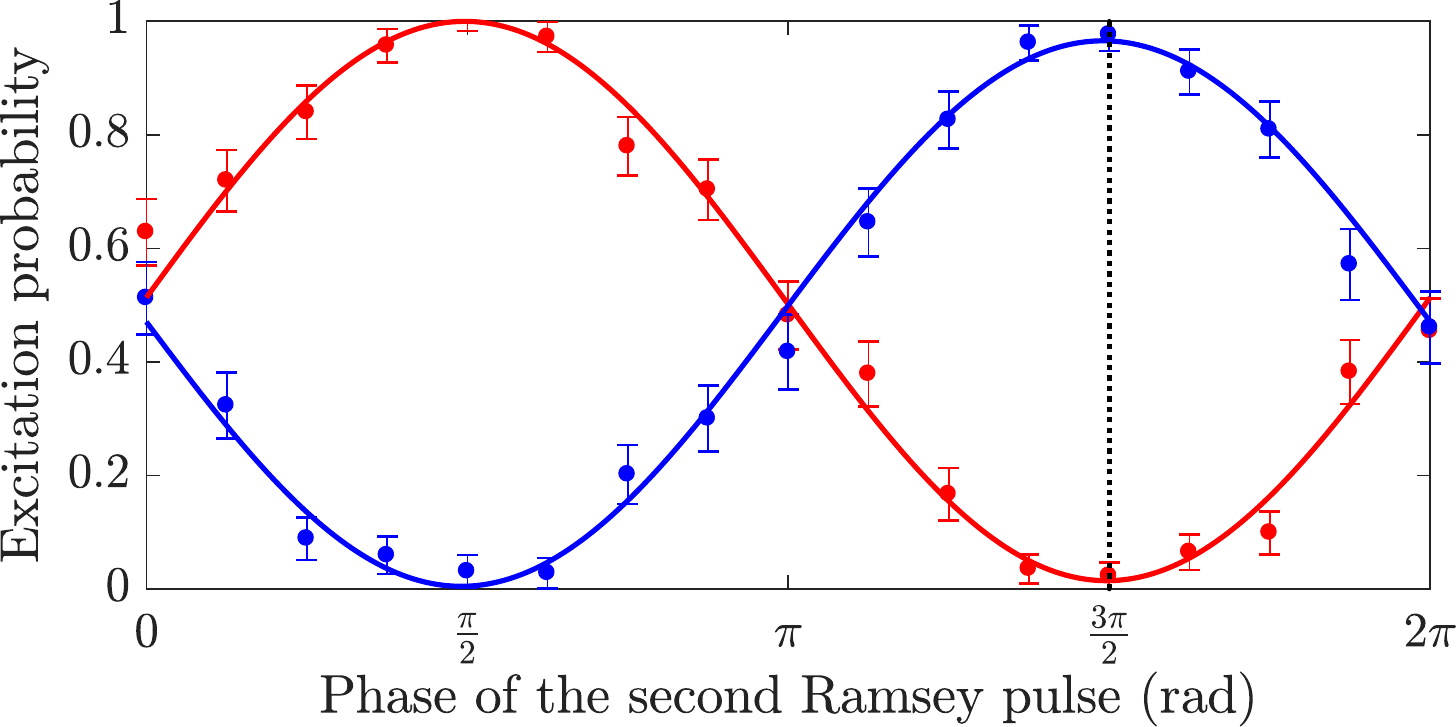}%
	\caption{\label{fig:PhaseShift} Ramsey measurement on the target qubit, when the AXY phase gate is applied to a two-ion system. The control qubit is initialized to $|0\rangle$ (red) or $|1\rangle$ (blue) to verify the conditional response of the gate. The former case results in a phase shift of $0.51(1)\pi$ with fringe contrast of $99^1_2\%$, the latter in a shift of $0.49(1)\pi$ and $96(2)\%$. Hence the phase difference between the two curves is $\pi$. The dashed line marks the phase setting for the second $\frac{\pi}{2}$ pulse, for which a CNOT operation results. Accordingly a phase of $\varphi = \frac{3\pi}{2}$ results in $|00\rangle \rightarrow |00\rangle$ for the red curve and $|10\rangle \rightarrow |11\rangle$ for the blue curve.}
\end{figure}

Figure \ref{fig:PhaseShift} shows the results obtained from using an AXY-16 sequence~(80 DD~pulses) designed for a $\Phi = \frac{\pi}{4}$ gate. Following Equation~\ref{eq:PhaseShift}, phase shifts $\Phi' = 0.51(1)\pi$ and $-0.49(1)\pi$ are measured with control qubit preparations $|0\rangle$ (red data points) and $|1\rangle$ (blue data points), respectively, with fringe contrast of $99^{+1}_{-2}\%$ and $96(2)\%$. 
Hence, a phase difference of $\pi$ is obtained between the red and blue curves at the target qubit when the control qubit's state is changed. The gate time is 6~ms.
Thus the phase shift matches the simulations, the conditional response of the sequence is confirmed and the fringe contrast allows for high-fidelity operations.

\subsection{Entanglement}
\begin{figure}[htb]
\centering
	\includegraphics[width= 0.8\columnwidth]{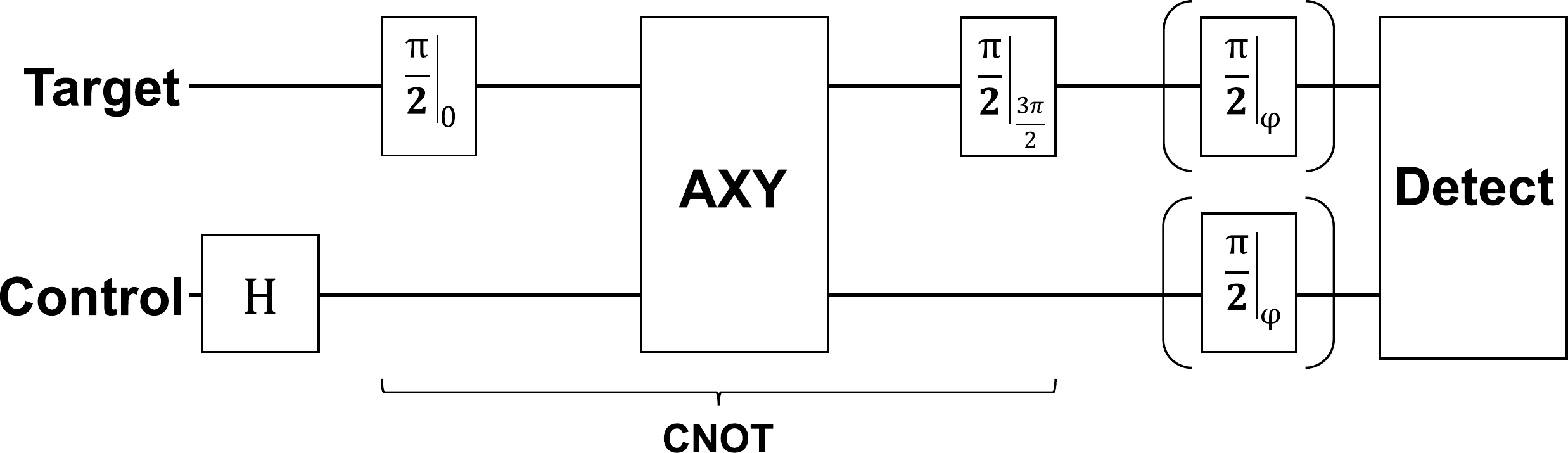}%
	\caption{\label{fig:BellSeq} RF sequence for a Bell state measurement. A Hadamard gate is applied to the control qubit to create an equal superposition state, then the AXY sequence framed by two $\frac{\pi}{2}$ pulses with phases $0$ and $\frac{3\pi}{2}$ creates a CNOT gate. Additional rotations to each qubit are optionally applied for rotating the measurement basis before state read-out.}
\end{figure}

Conditional phase gates are applicable for the generation of entanglement. For instance, a conditional $\Phi = \frac{\pi}{4}$ phase gate allows for a CNOT operation and hence preparation of maximally entangled Bell states. The dashed vertical line in Fig.~\ref{fig:PhaseShift} marks the phase setting for the second $\frac{\pi}{2}$ pulse, for which a CNOT operation is achieved. This pulse must be executed accordingly with a phase of $\varphi = \frac{3\pi}{2}$ to receive a flip of the target qubit if the control qubit is in state $|1\rangle$. 
The maximally entangled Bell state $|\Phi\rangle = \frac{|00\rangle + |11\rangle}{\sqrt{2}}$ is produced by adding a (single-qubit) Hadamard operation only, as shown in Fig.~\ref{fig:BellSeq}. 
Thus, the control qubit is initialized into an equal superposition state before the CNOT gate is applied.

\begin{figure}[htb]
\centering
    \begin{subfigure}[b]{0.48\columnwidth}
    \includegraphics[width = \columnwidth]{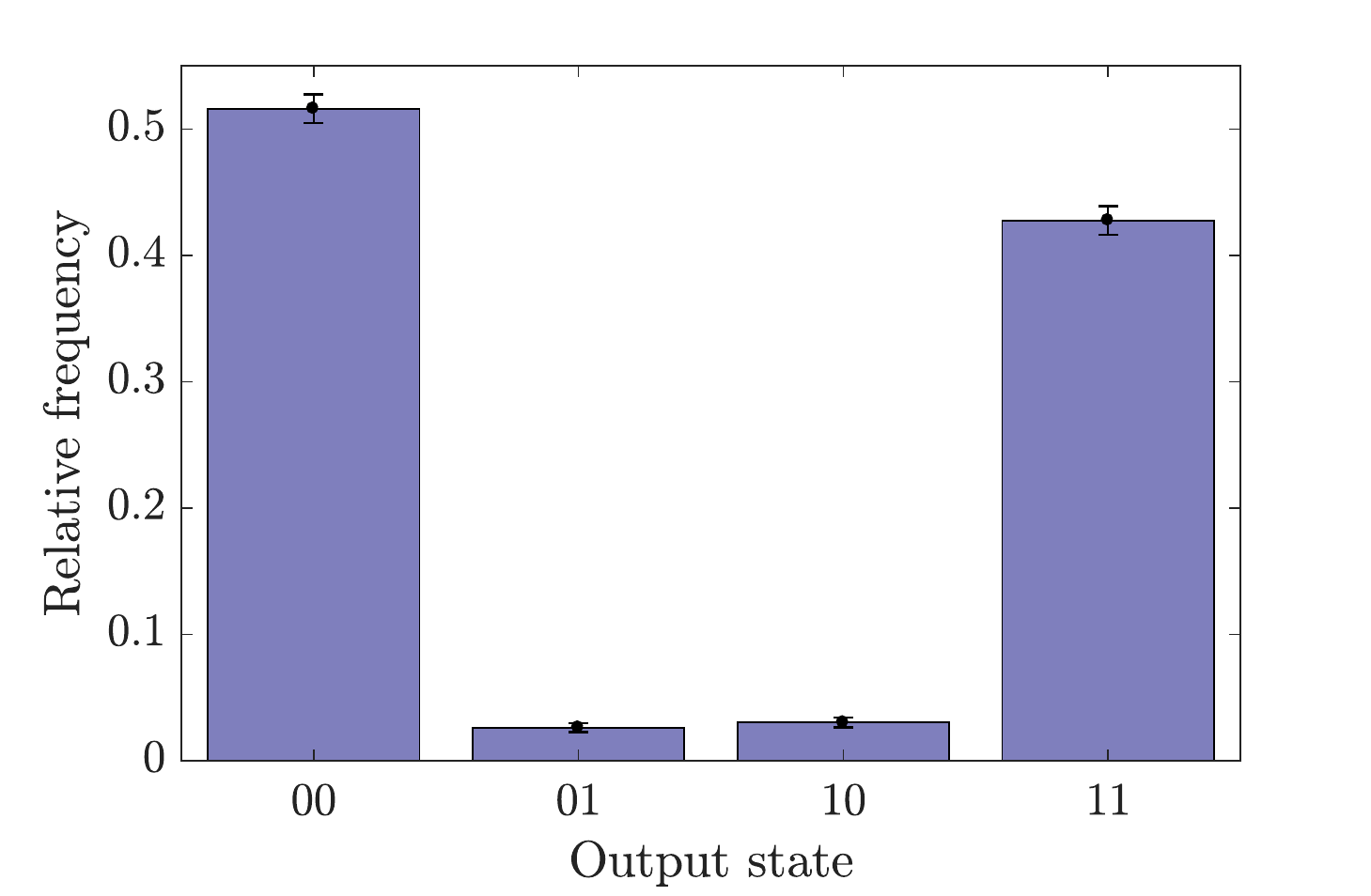}%
    \caption{}
    \end{subfigure}
\hfill
    \begin{subfigure}[b]{0.48\columnwidth}
    \includegraphics[width = \columnwidth]{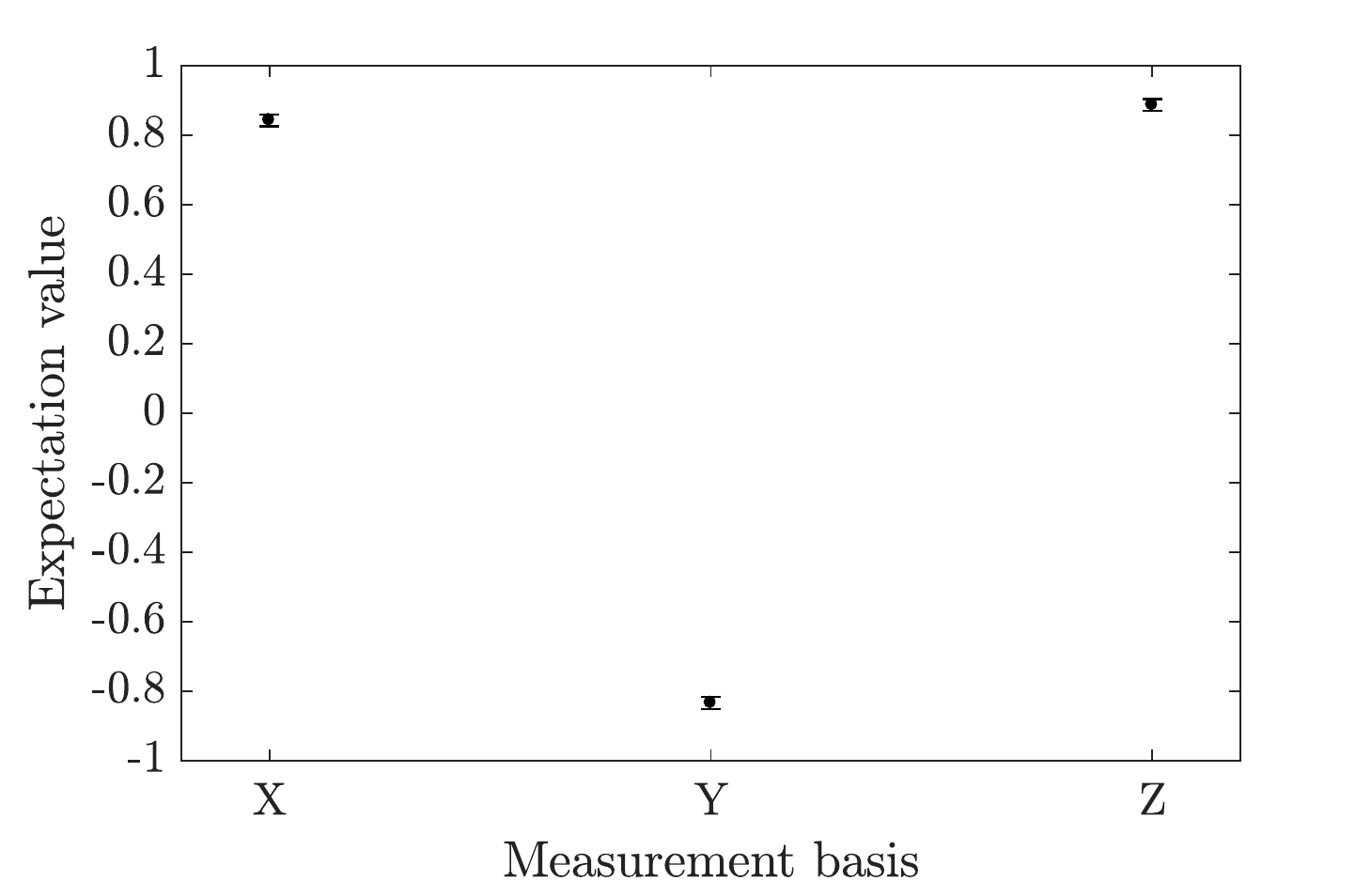}%
    \caption{}
    \end{subfigure}
\caption{\label{fig:Bell3Pt}Results of a Bell state measurement of the state $|\Phi\rangle = \frac{1}{2}(|00\rangle + |11\rangle)$ created using the AXY gate sequence. (a): Relative frequencies of the four basis states $\{|00\rangle, |01\rangle,|10\rangle,|11\rangle\}$, measured in the z-basis. (b): Expectation values for the x, y and z basis, derived from 3 similar measurement steps, when the basis is rotated before state read-out. From these, the Bell state fidelity of 89.1(1.5)\% is calculated.}
\end{figure}

An entangled state shows quantum correlations when measured in different bases. As state detection is a projective measurement to the z-basis only, an additional set of RF-pulses is appended to the pulse sequence to allow for measurements in the x and y basis as well. These RF pulses map the respective populations to the z basis before shining in the laser. The Bell state fidelity is then reconstructed from the correlations in the x, y and z basis, respectively.
Often a full parity scan is used to asses correlations in the different bases, by scanning the phase of the basis-changing pulses from $0$ to $2\pi$~\cite{Sackett2000}.
We reduce measurement efforts considerably by probing only at two specific rotations (in addition to the z-measurement itself, without the pulses). In particular, settings of $\varphi = \{\frac{\pi}{4}, \frac{3\pi}{4}\}$ correspond to the x and y basis (the extreme points of the parity curve). Using these measurement results, the Bell state fidelity is determined according to~\cite{Terhal2002, Guehne2003} as
\begin{equation}
  F = \frac{\langle\sigma_x\otimes\sigma_x\rangle - \langle\sigma_y\otimes\sigma_y\rangle + \langle\sigma_z\otimes\sigma_z\rangle + 1}{4},
 \label{eq:Bell}
\end{equation}
with the expectation values given by
\begin{equation}
\langle\sigma_i\otimes\sigma_i\rangle = P_{00}^i + P_{11}^i - \left(P_{10}^i + P_{01}^i\right),
\label{eq:ExpValues}
\end{equation}
for $i = \{x,y,z\}$ and measured relative frequencies $P^i_{jk}$ of each of the four basis states in the respective basis. Figure \ref{fig:Bell3Pt} depicts the results of these measurements, showing the relative frequencies in the z-basis on the left and the expectation values in all three bases on the right. According to Equation~\ref{eq:Bell}, a Bell state fidelity of 89.1(1.5)\% is achieved in this experiment.

Furthermore, avoiding full quantum state tomography, we can also provide a lower bound on the logarithmic negativity which constitutes a computable measure of entanglement~\cite{Plenio2005}. The logarithmic negativity of a density operator $\rho$ is defined as $E_N(\rho) = \log_2 tr|\rho^{\Gamma}|$ where $\rho^{\Gamma}$ denotes the partial transposition of $\rho$. As its exact determination would require full state tomography we follow~\cite{Audenaert2006} and obtain a very good lower bound based on the measurement of the correlators $\langle \sigma_x\otimes\sigma_x\rangle$, $\langle \sigma_y\otimes\sigma_y\rangle$ and $\langle \sigma_z\otimes\sigma_z\rangle$ and the determination of density matrix with the least logarithmic negativity that is compatible with the measured correlators. This minimization can be carried out analytically and yields
\begin{equation}
	E_N > \rm log_2^+\left(\frac{(1+ |\langle\sigma_x\otimes\sigma_x\rangle| + |\langle\sigma_y\otimes\sigma_y\rangle| + |\langle\sigma_z\otimes\sigma_z\rangle|)}{2}\right),
\end{equation}
with $\rm log_2^+(x) = \rm max(0, \rm log_2(x))$.  From the measurement data of Fig.~\ref{fig:Bell3Pt}, we obtain
\begin{equation}
	E_N > 0.832(17).
\end{equation}

\begin{figure}[ht]
\centering
\includegraphics[width = 0.5\columnwidth]{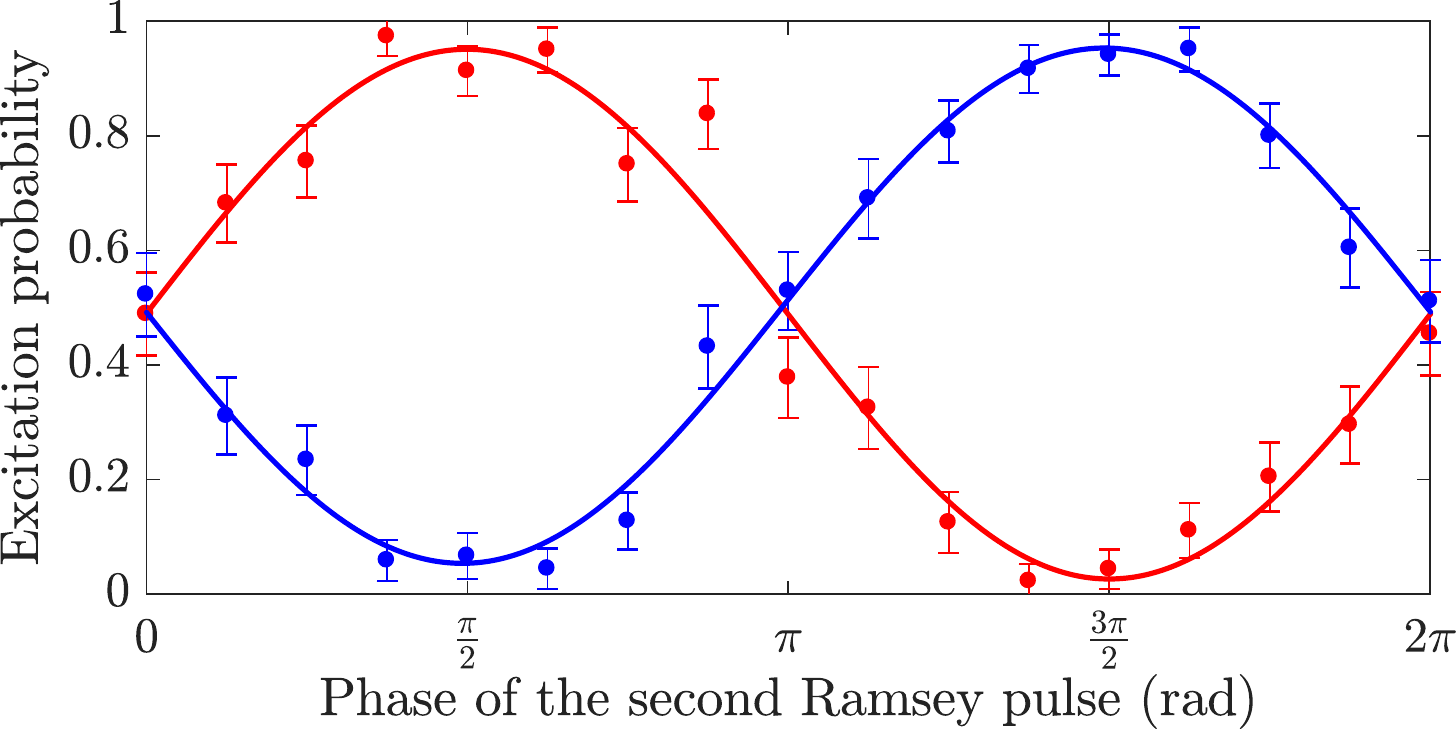}%
\caption{\label{fig:BellRamseyComp}Result of a Ramsey measurement on the target qubit, when the AXY phase gate is applied to a two ion system. The control qubit is initialized to $|0\rangle$ (red) or $|1\rangle$ (blue), resulting in conditional phase shifts of $0.50(1)\pi$ with fringe contrast of $92(4)\%$ and $-0.49(1)\pi$ with $89(3)\%$ respectively. The reduction of fringe contrast compared to the previous measurements (Fig.~\ref{fig:PhaseShift}) matches with the observed Bell fidelity of 89.1(1.5)\% and is attributed to an instability of the current experimental setup.}
\end{figure}

It should be remarked that these results do not match the expected fidelity based on the previously reported Ramsey fringe contrast~(Fig.~\ref{fig:PhaseShift}). This discrepancy is due to fluctuations in the current experimental setup - an unstable mechanical or electric connection within the vacuum chamber is suspected - preventing stable results with highest fidelities. However, it has been observed that such instabilities affect the measurements similarly. Back-to-back Ramsey and Bell measurements always yield compatible results in terms of fringe contrast and Bell fidelity, as indicated by Fig.~\ref{fig:BellRamseyComp}.
This is to be expected, since the two measurement sequences (two-ion Ramsey and Bell state) differ only by single-qubit rotations, which have low error rates \cite{Piltz2014}.
Hence, the achievable Bell fidelity is expected to be limited mainly by the Ramsey fringe contrast, which translates into the errors of the relative frequencies of Eq.~\ref{eq:ExpValues} for correctly measuring the target qubit.
It is therefore implied by the high-fidelity Ramsey results as shown in Fig.~\ref{fig:AXY_Ramsey} that a Bell fidelity of 99\% is reachable, even though it has not been independently measured yet.

\section{\label{sec:Robustness}Robustness}
We investigate in detail the resilience of the gate to several potential sources of error: the excitation of the ions' COM vibrational mode, pulse errors, and timing mismatch. Using Ramsey measurements as before, fringe contrast and phase shift serve as figures of merit for the gate performance. All of the following measurements are performed with and normalized to additional reference measurements.

Decoherence is a major obstacle for quantum logic operations.
In the current experimental setup, the gate time of 6~ms faces a coherence time of $<500~\mu$s when no DD is applied. 
However, using pulsed DD preserves coherence for even longer gate times. Quantum logic operations with gate times of 30~ms have been successfully implemented~\cite{Sriarunothai_2018} using this method. In case of the AXY gate, sequences of 80 or 160 pulses (AXY-16 and AXY-32 respectively) show the best results in terms of balancing decoupling and pulse errors. Increasing the number of pulses further has no benefit in gate fidelity, since the pulse errors start to become detrimental.

\subsection{Vibrational excitation}
\begin{figure}[htb]
\centering
    \begin{subfigure}[b]{0.48\columnwidth}
	\includegraphics[width = \columnwidth]{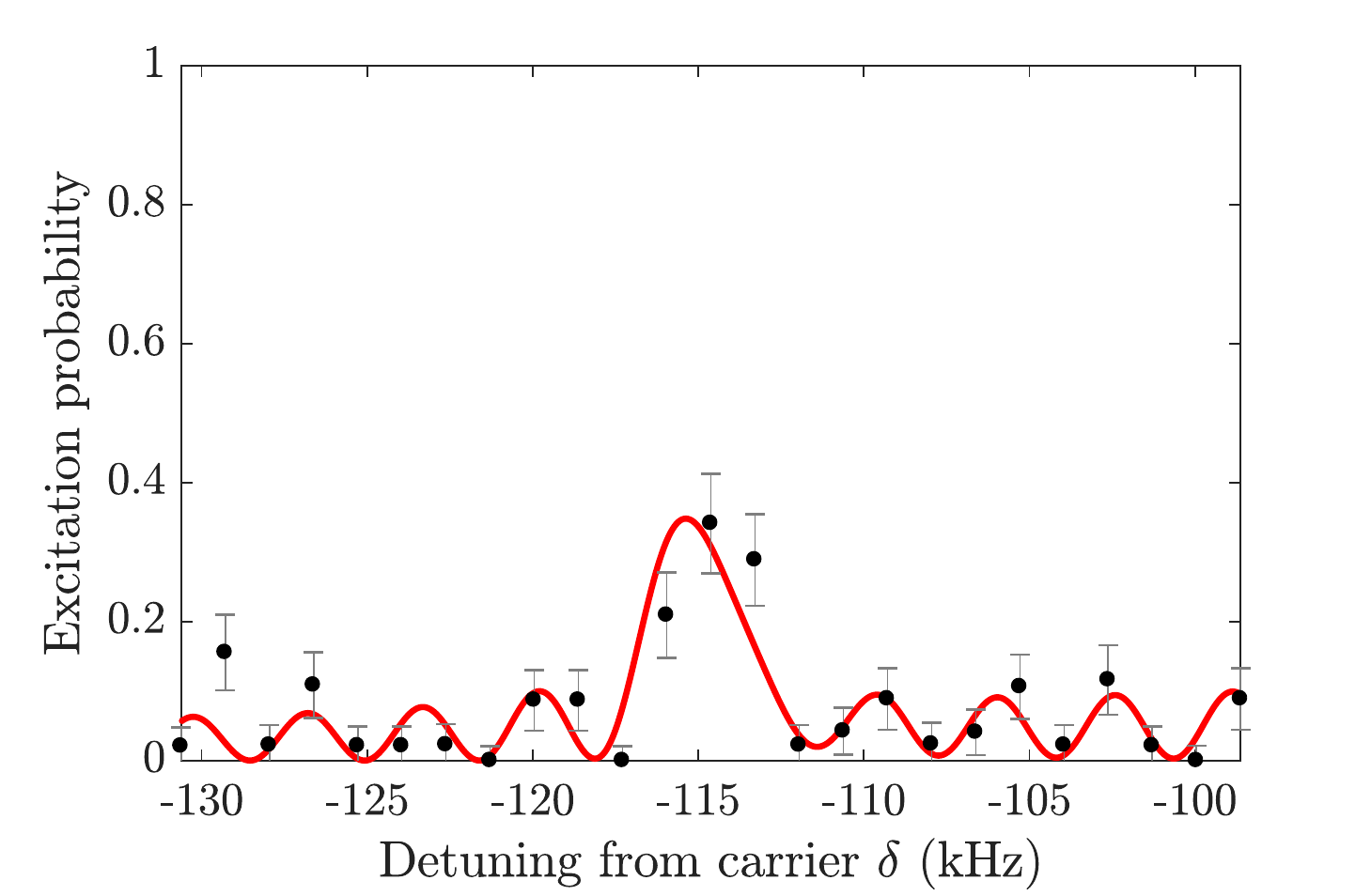}%
    \caption{}
    \end{subfigure}
\hfill
    \begin{subfigure}[b]{0.48\columnwidth}
	\includegraphics[width = \columnwidth]{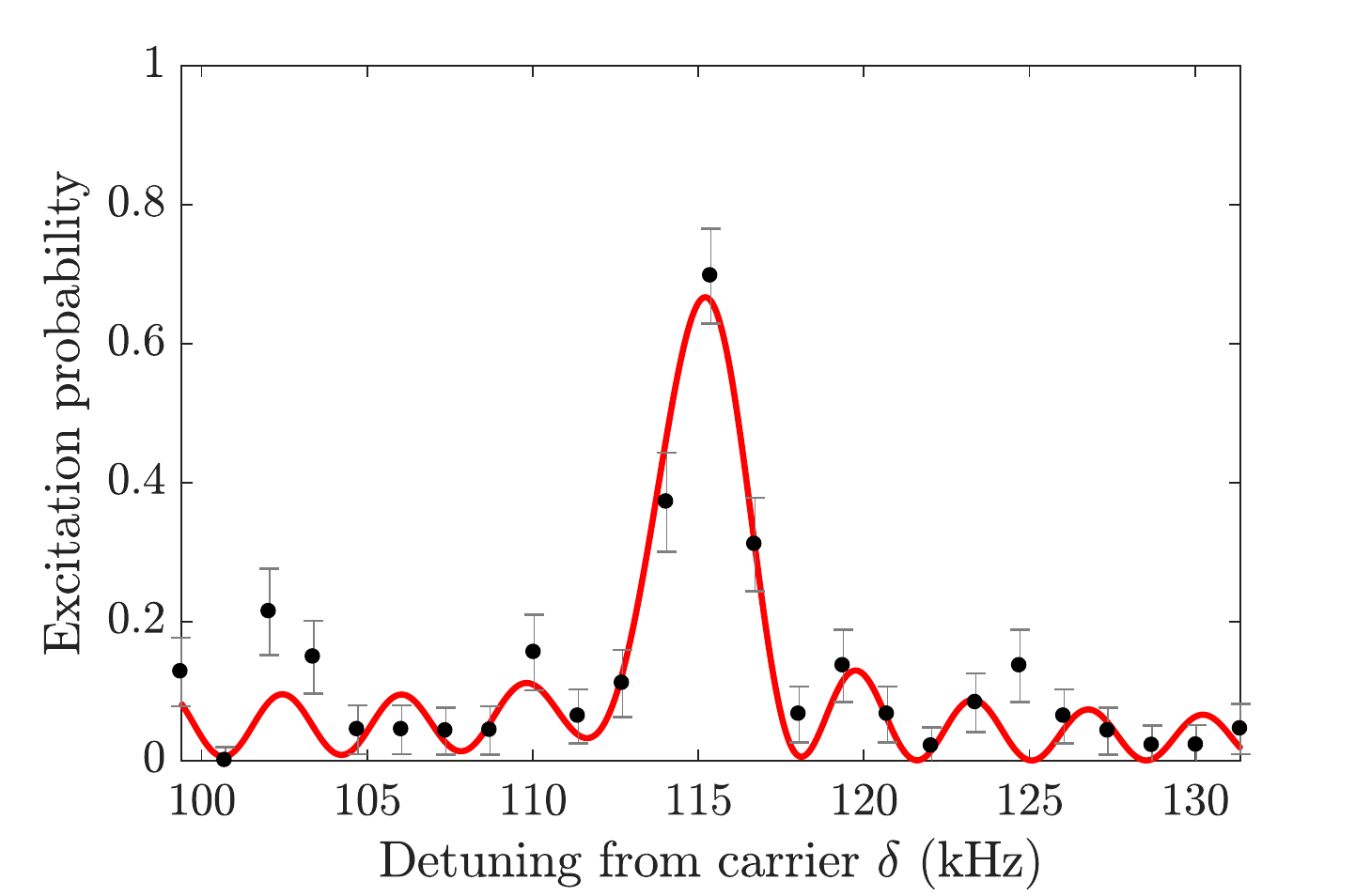}%
	\caption{}
    \end{subfigure}
\caption{\label{fig:SB_Spectroscopy}Measurement of the mean COM vibrational mode excitation $\bar{n}$  by sideband spectroscopy. A RF pulse with variable frequency is applied to excite motional sidebands of the qubit resonance. The asymmetry of excitation of the red (a) and blue (b) sideband is used to calculate $\bar{n}$~\cite{Sriarunothai2018}. The red curve displays a fit for both of the sidebands, which results in $\bar{n}=1.0(5)$~phonons.}
\end{figure}

Many conditional gates rely on ground state cooling of an ion chain. We investigate the effect of initial COM mode excitation on the gate performance. Different initial thermal excitations with mean phonon number $\bar{n}$ between 1.0(5) up to the Doppler cooling level, estimated to the order of 100~phonons, are achieved by placing a variable heating time after initial sideband cooling, before the beginning of RF manipulation. Given the heating rate of our setup of $0.12$~phonons/ms, the mean phonon number can be adjusted by varying the heating time accordingly. 
Each phonon setting is verified by sideband spectroscopy measurements, consisting of a single RF pulse with fixed time of 300~$\mu$s, whose frequency is scanned around the red and blue sidebands of the qubit resonance, shifted by $\nu_{1}$ below and above the atomic carrier transition respectively. The mean phonon number is obtained by fitting the probability to populate state $|1\rangle$ as 
\begin{equation}
	P_{|1\rangle}(\delta, \bar{n}) = P_C(\delta, \bar{n}) + \sum_{k = 1}^2\left[P_{RSB}^{(k)}(\delta, \bar{n}) + P_{BSB}^{(k)}(\delta, \bar{n})\right],
\end{equation}
taking into account the carrier~($P_C$) as well as the first- and second-order red~($P_{\rm RSB}$) and blue~($P_{\rm BSB}$) sidebands at given detuning~$\delta$ and mean phonon number $\bar{n}$~\cite{Sriarunothai2018}.
Figure~\ref{fig:SB_Spectroscopy} displays the result of such measurement with $\bar{n} = 1.0(5)$ as the red sideband is being suppressed compared to the blue one.

\begin{figure}[htb]
\centering
    \begin{subfigure}[b]{0.48\columnwidth}
    \includegraphics[width = \columnwidth]{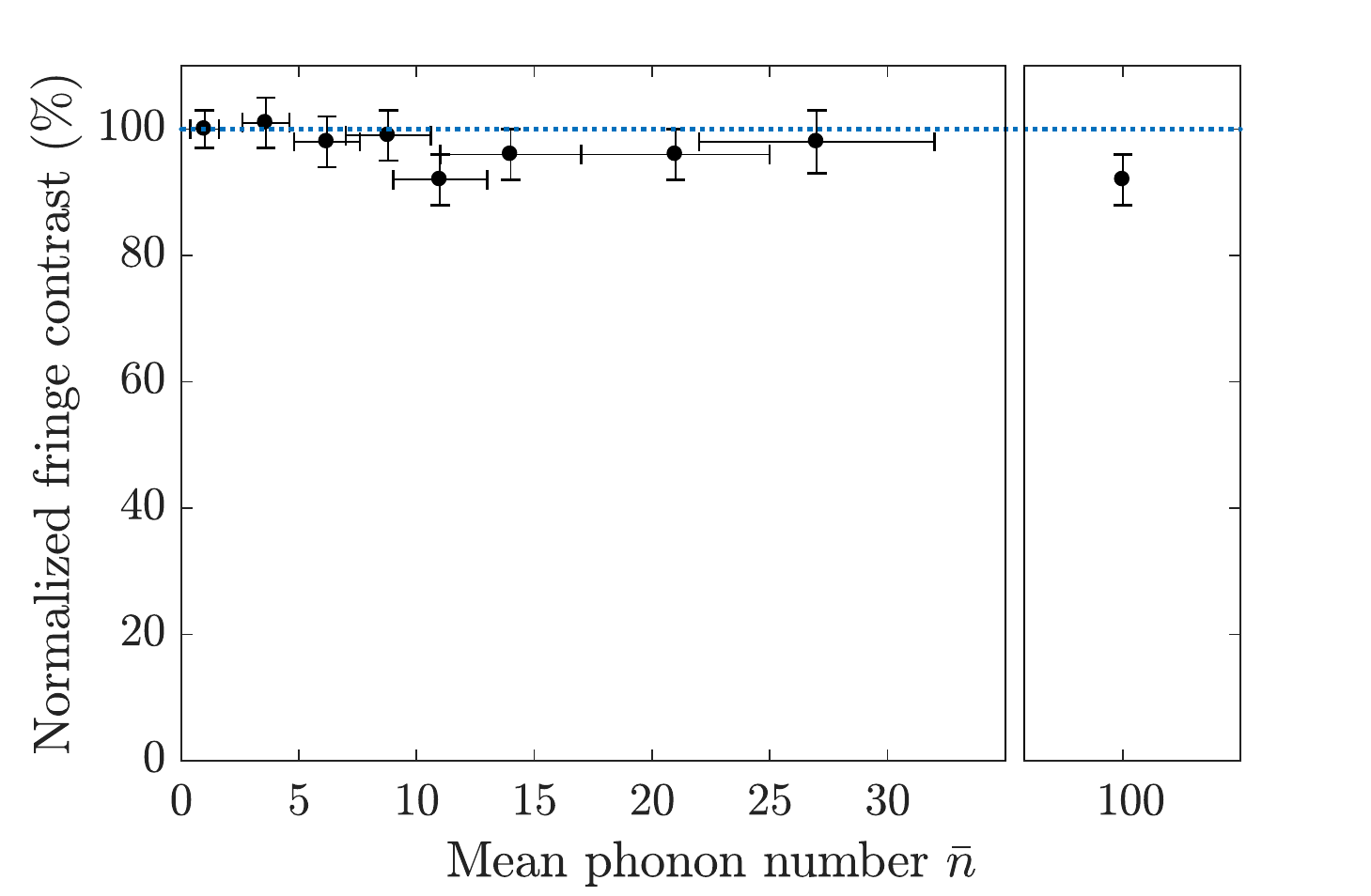}%
    \caption{}
    \end{subfigure}
\hfill
    \begin{subfigure}[b]{0.48\columnwidth}
    \includegraphics[width = \columnwidth]{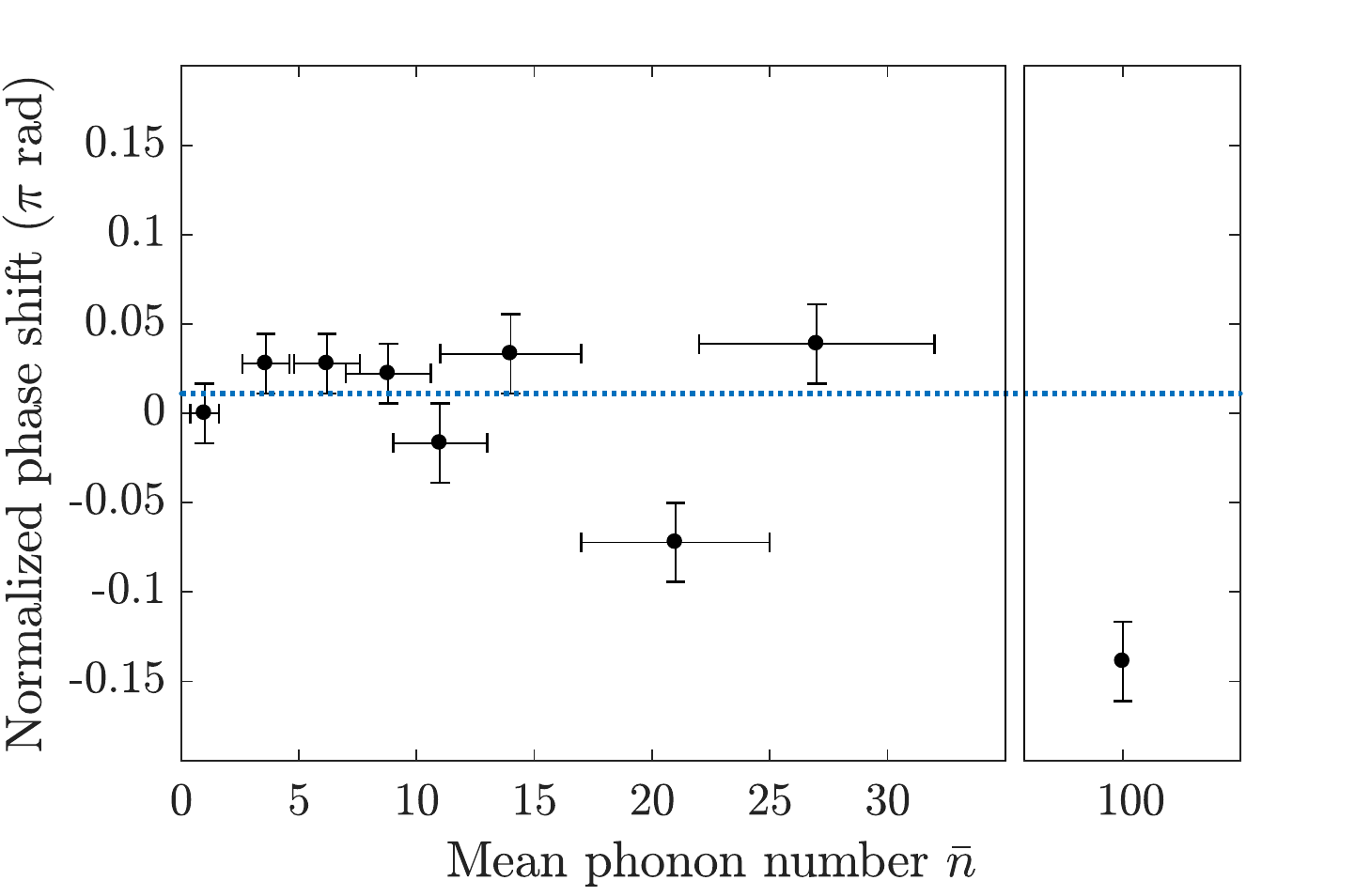}%
	\caption{}
    \end{subfigure}
\caption{\label{fig:Temperature}Influence of mean thermal phonon number $\bar{n}$ of the COM vibrational mode on the resulting Ramsey fringe contrast (a) and phase shift (b) of the AXY gate. An AXY-16 sequence is applied to generate a $\frac{\pi}{4}$ phase gate, while $\bar{n}$ is varied. This is achieved by a variable heating time before the gate sequence, after RF sideband cooling~(SBC) close to the motional ground state. The last point depicts the Doppler cooled state, when no SBC is applied at all. The dashed line indicates the reference setting of 1.0(5)~phonons, to which all results are normalized.}
\end{figure}

The effect of varying the initial COM mode excitation on the phase gate can be seen in Fig.~\ref{fig:Temperature}. There is no significant decrease in fringe contrast nor a change in the achieved phase shift up to $\bar{n} \approx 10$ phonons as compared to an ion crystal cooled close to the ground state. While the resulting phase starts to deviate towards higher $\bar{n}$, the contrast is still stable with values \textgreater90\% even when no sideband cooling is applied. 

\subsection{Secular axial trap frequency}
\begin{figure}[htb]
\centering
	\includegraphics[width = 0.45\columnwidth]{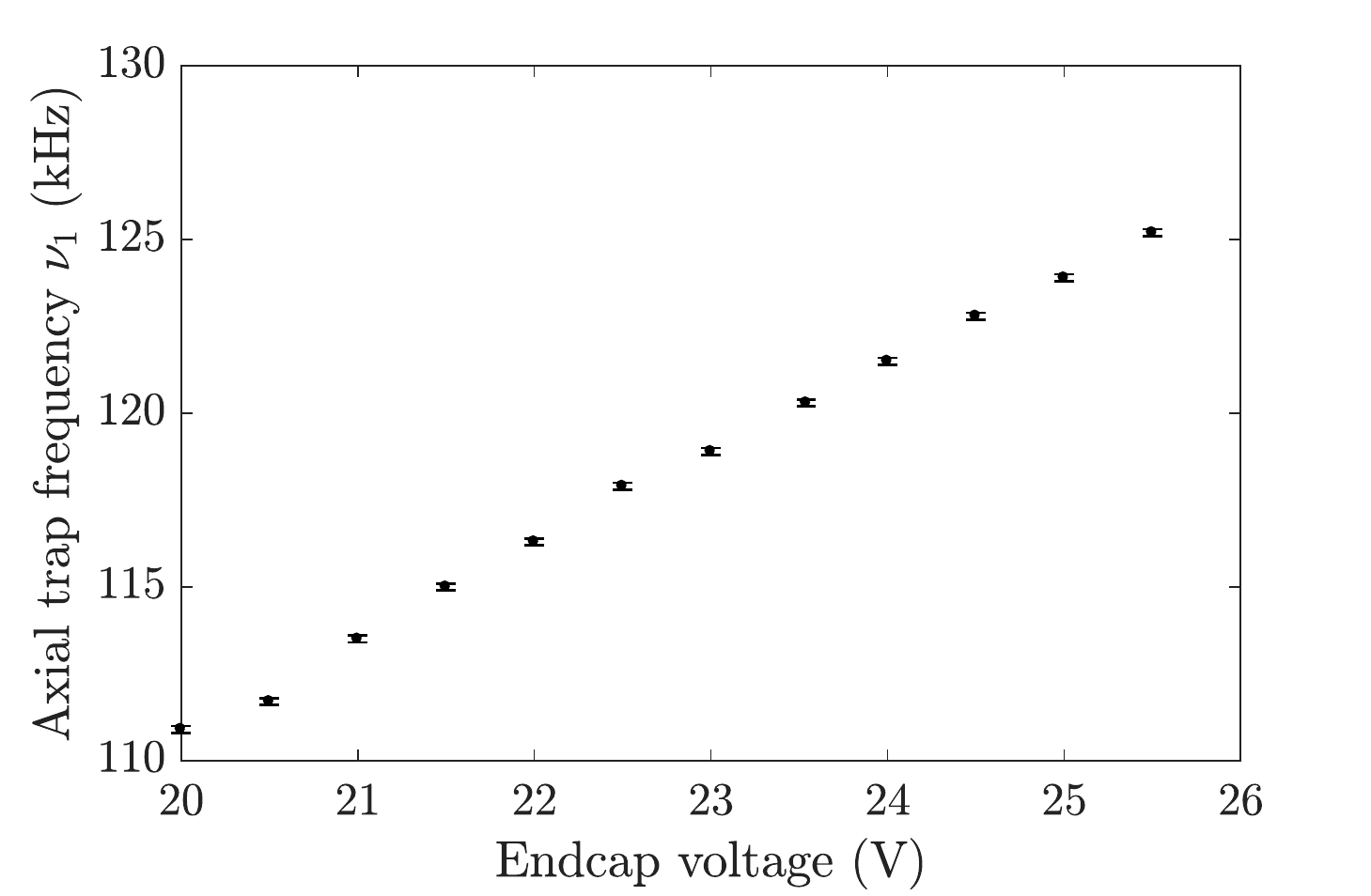}%
	\caption{\label{fig:NuVsEndcaps} Measurement of the trap frequency. Changing the voltage applied to the endcap electrodes, the axial trapping potential and hence axial motional frequencies are varied. A tickling method using a small AC voltage to excite the ion motion is used to determine the respective trap frequencies. }
\end{figure}

\begin{figure}[htb]
\centering
    \begin{subfigure}[b]{0.48\columnwidth}
    \includegraphics[width = \columnwidth]{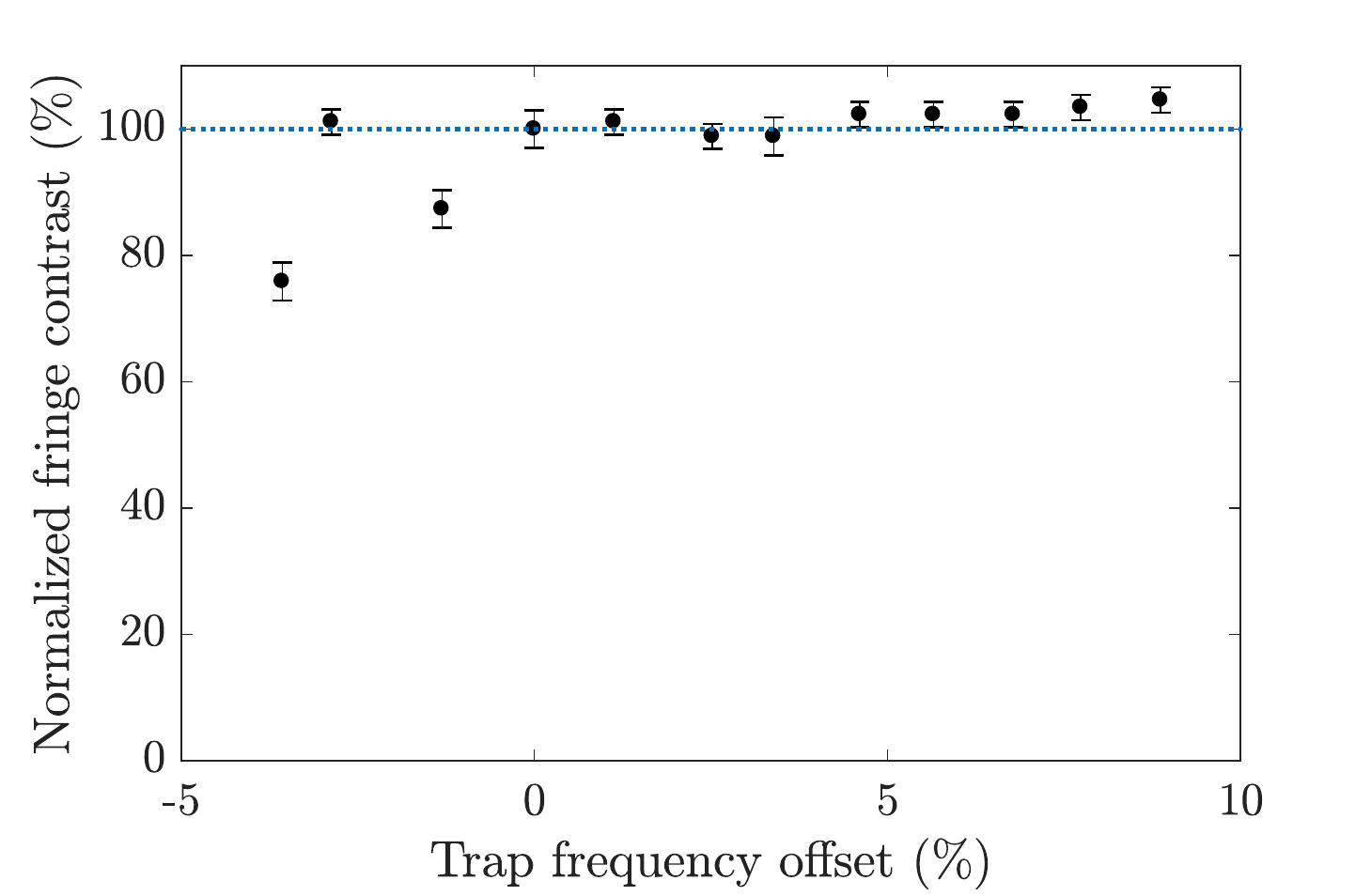}%
    \caption{}
    \end{subfigure}
\hfill
    \begin{subfigure}[b]{0.48\columnwidth}
    \includegraphics[width = \columnwidth]{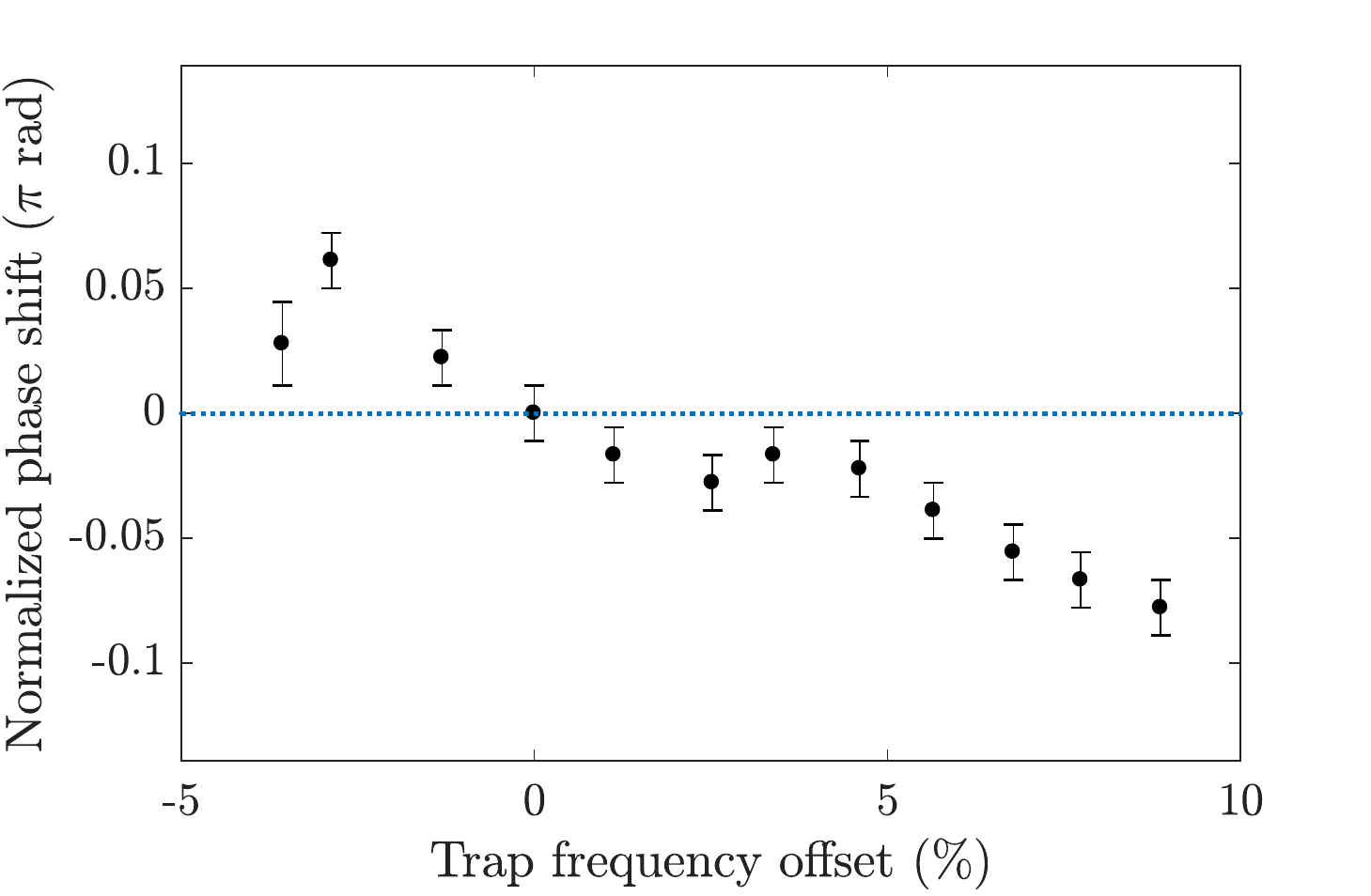}%
	\caption{}
    \end{subfigure}
\caption{\label{fig:TrapFreq}Effect of the trap frequency on the resulting Ramsey fringe contrast (a) and phase shift (b) of the AXY gate. The axial trap frequency is varied, deliberately inducing a mismatch between the pulse timings of the AXY sequence and the motional frequencies. An AXY-16 sequence is applied to generate a $\frac{\pi}{4}$ phase gate, the dashed line indicates the reference setting of $\nu_t = 2\pi\times115$~kHz, to which all results are normalized.}
\end{figure}

The gate mechanism is based on tuning pulse timings of the AXY sequence with respect to the motional mode frequencies of the ion crystal. Changing the endcap voltages, the axial trapping potential and thus the mode frequencies can be detuned with respect to the reference setting, based on which the sequence timings are calculated. 
The respective secular trap frequency is measured by a method called tickling. In this method a small alternating voltage is applied to one of three available compensation electrodes around the trap center. When its frequency matches the secular frequency of the COM ion motion, the latter is excited and the crystal starts to melt. With this, we typically achieve a resolution of $\leq10$~Hz for the axial trap frequency~$\nu_1$. Figure~\ref{fig:NuVsEndcaps} displays endcap voltage settings and the measured axial COM trap frequencies. 
Based on this calibration, Fig.~\ref{fig:TrapFreq} displays the effect of mismatches of the theoretically assumed and the actual axial trap frequency on the performance of the phase gate. While the contrast of the gate is not significantly affected, accurate setting of the mode frequency is crucial to match the pre-computed phase shift. Given the resolution of $<10$~Hz achieved with the tickling method, knowledge of the mode frequencies to the $<10^{-4}$ level is provided. As suggested from Fig.~\ref{fig:TrapFreq}, this linearly translates into the phase error of the gate, i.e. infidelities below $10^{-4}$. Low pass filters with cut-off frequencies $<1$~Hz are used to stabilize the endcap voltages, i.e. errors due to fluctuations are suppressed to below $10^{-5}$. Towards the lower endcap voltages jumps occur even in fringe contrast, indicating instabilities of the trapping potential.

\subsection{Pulse errors and sequence timing}
\begin{figure}[htb]
\centering
    \begin{subfigure}[b]{0.48\columnwidth}
	\includegraphics[width = \columnwidth]{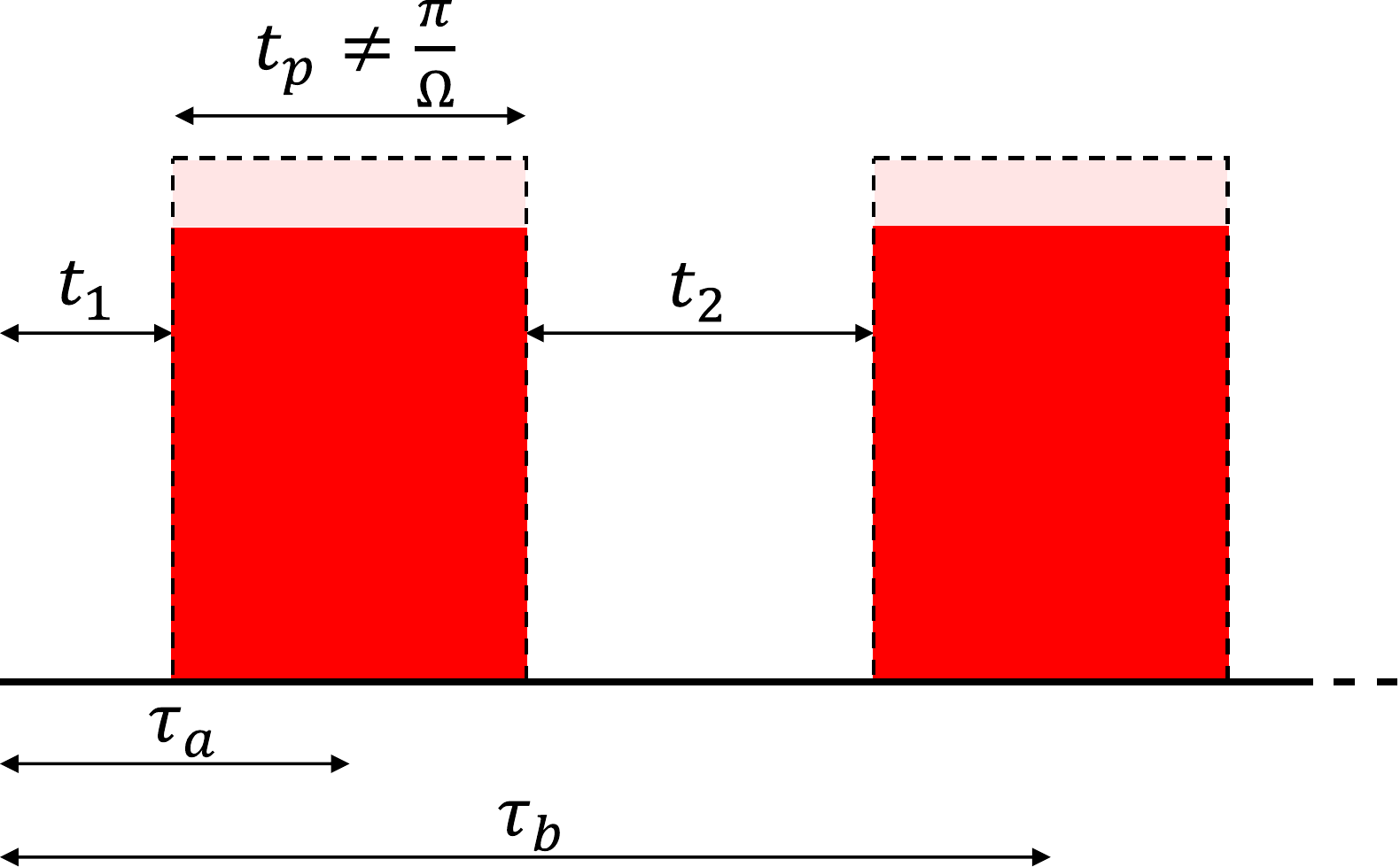}%
    \caption{}
    \end{subfigure}
\hfill
    \begin{subfigure}[b]{0.48\columnwidth}
	\includegraphics[width = \columnwidth]{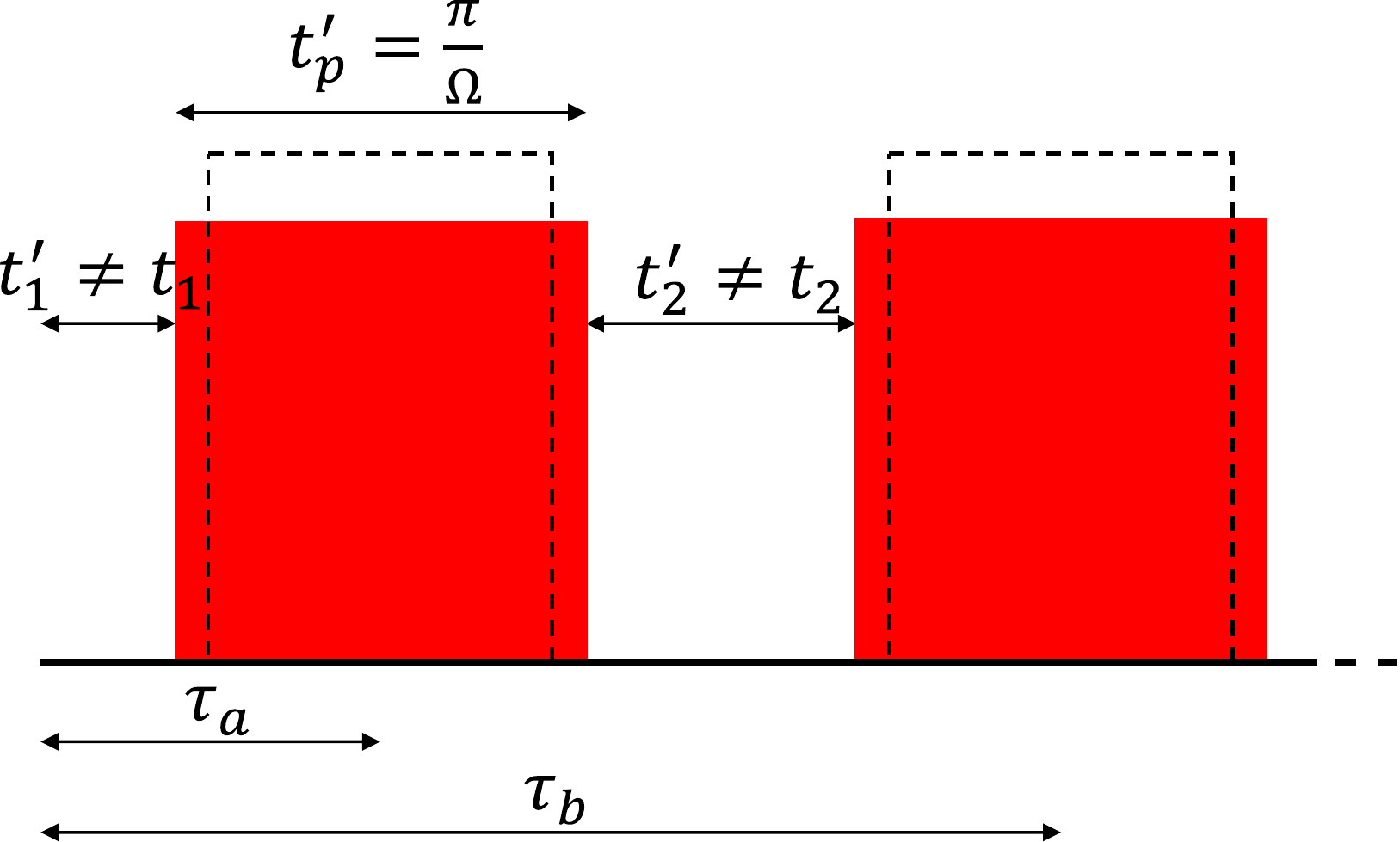}%
	\caption{}
    \end{subfigure}
\caption{\label{fig:PulseErrors}Two types of errors induced by instabilities of the RF power. The first two pulses of an AXY block are shown as an example, where the dashed lines indicate the calculated pulse width and amplitude, respectively, while the red areas sketch the actually applied pulses. (a): Pulse area errors cause all pulses to differ from $\pi$ rotations, as the pulse time does not match $t_p = \frac{\pi}{\Omega}$. Timings of the AXY sequence are still correct. (b): Re-calibration of the RF amplitudes allows for correct $\pi$ rotations, at the cost of a different pulse time $t_p'$. This causes spacings $\{t_1,t_2,...\}$ in between pulses to differ from the calculated sequence. Center positions of each pulse defined by the sequence parameters $\tau_a$ and $\tau_b$ are not affected.}
\end{figure}

Instability of the RF power is another common source of error, since it translates directly into the Rabi frequencies of the qubit system.
Here, we distinguish between two possible types of errors, which can be caused by this, as sketched in Fig.~\ref{fig:PulseErrors}.
First, we investigate pulse area errors due to a mismatch between the theoretical Rabi frequency, used to calculate the pulse widths and the experimental Rabi frequency defined by the applied RF power. This causes all pulses to differ from $\pi$ rotations, while sequence timings still match the calculations. Such errors can, for instance, be caused by slow drifts of the RF power due to temperature changes of the electronics.

\begin{figure}[htb]
\centering
    \begin{subfigure}[b]{0.48\columnwidth}
	\includegraphics[width = \columnwidth]{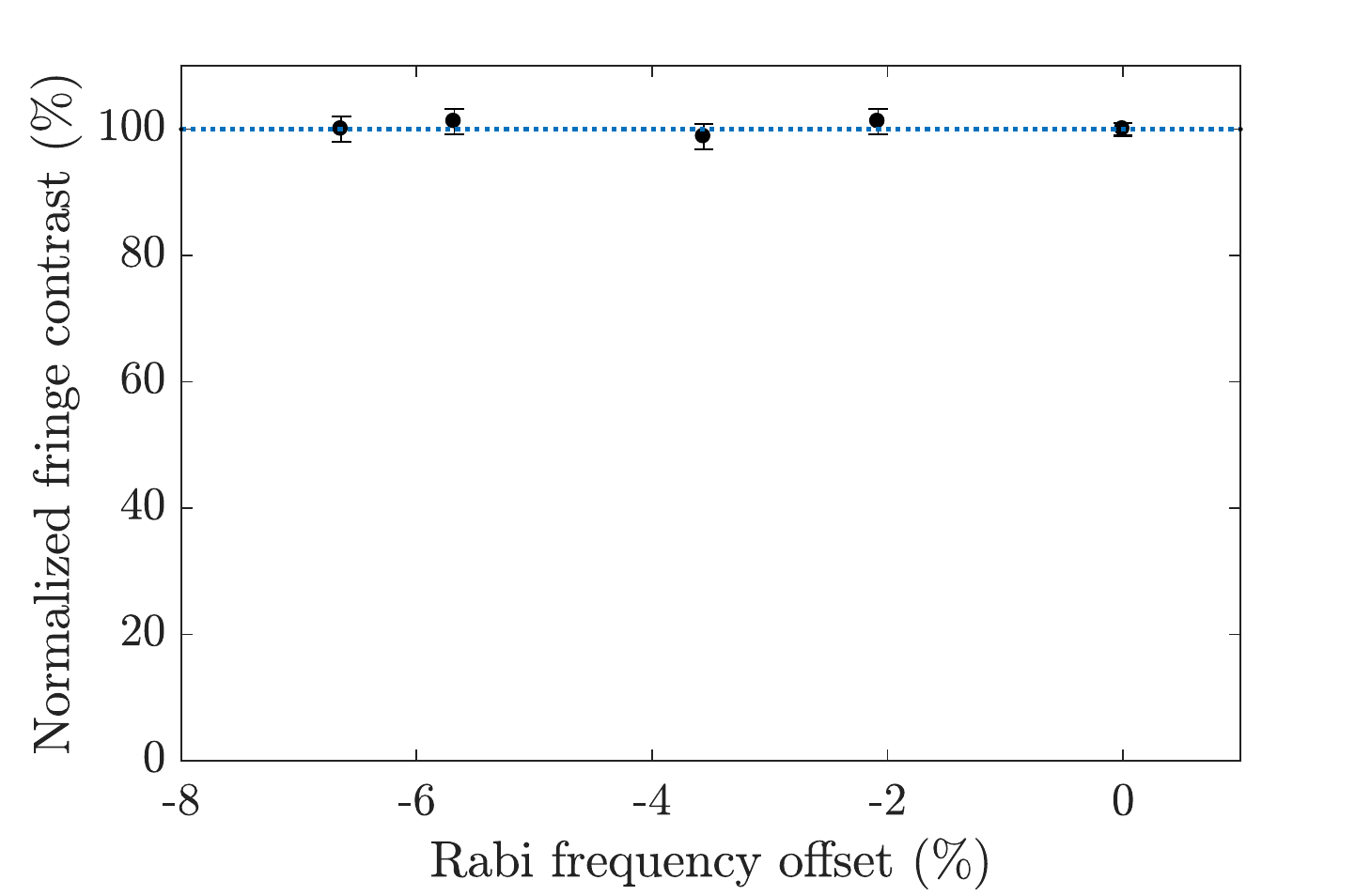}%
	\caption{}
    \end{subfigure}
\hfill
    \begin{subfigure}[b]{0.48\columnwidth}
	\includegraphics[width = \columnwidth]{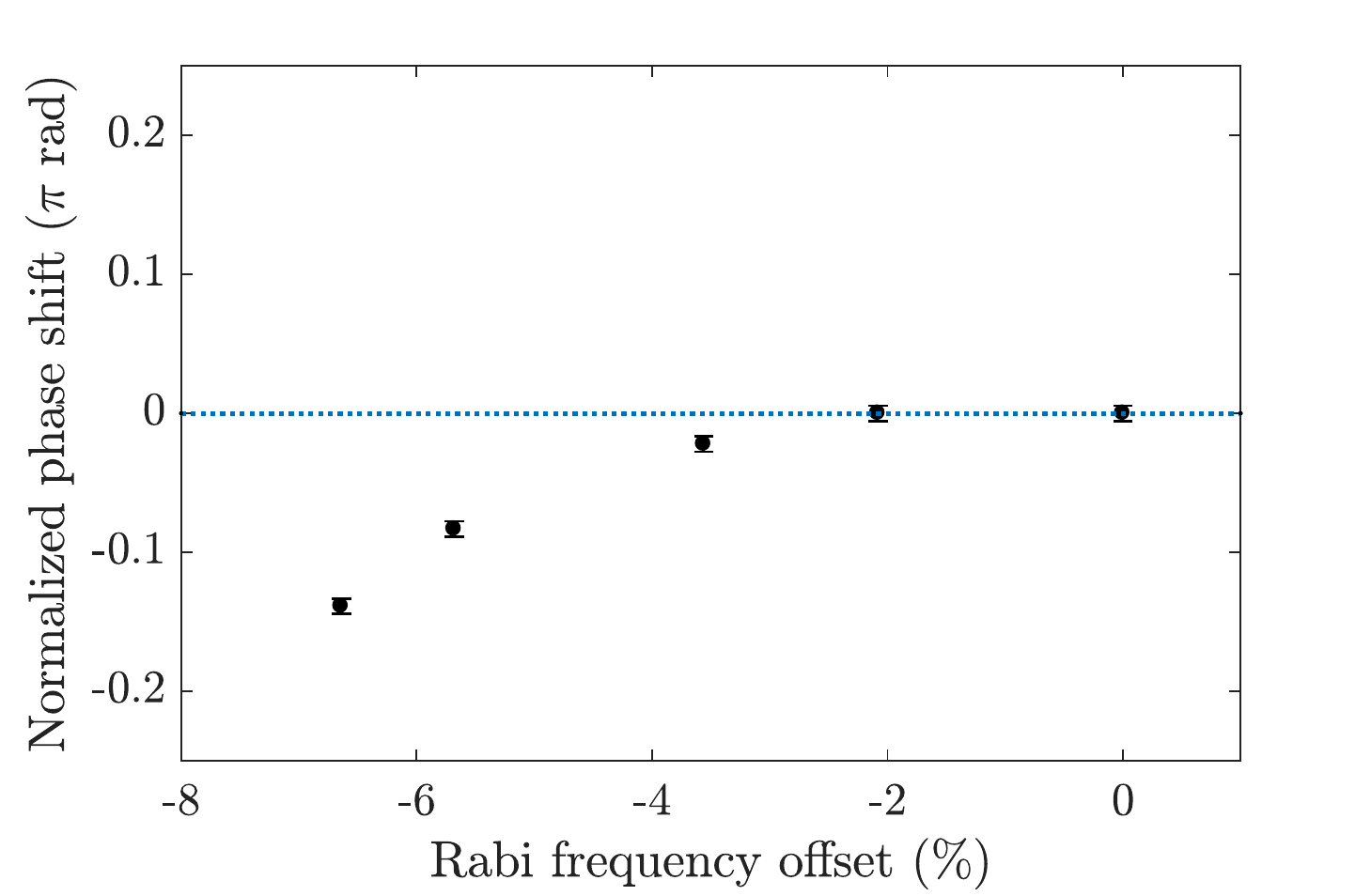}%
	\caption{}
    \end{subfigure}
	\caption{\label{fig:OmegaError}Effect of pulse errors on the resulting Ramsey fringe contrast (a) and phase shift (b) of the AXY gate. An AXY-16 sequence is applied to generate a $\frac{\pi}{4}$ phase gate, with an offset of the pulse times with respect to the ions' Rabi frequencies. Results are normalized to the reference setting of zero offset as indicated by the dashed line.}
\end{figure}

Figure~\ref{fig:OmegaError} shows the resulting contrast and phase of an AXY-16 gate, if a mismatch is deliberately introduced to the RF power by changing the amplitude setting of the signal generator. For each point, the offset is confirmed by a separate measurement of the actual Rabi frequency.
On the one hand, these results confirm the robustness of the DD pulses against pulse errors, since there is no significant effect on the measured fringe contrast. On the other hand, the phase shift is affected for mismatches $>2\%$. However, keeping RF amplitudes stable to less than $1\%$ is well within current experimental limits. 

\begin{figure}[htb]
\centering
    \begin{subfigure}[b]{0.48\columnwidth}
    \includegraphics[width = \columnwidth]{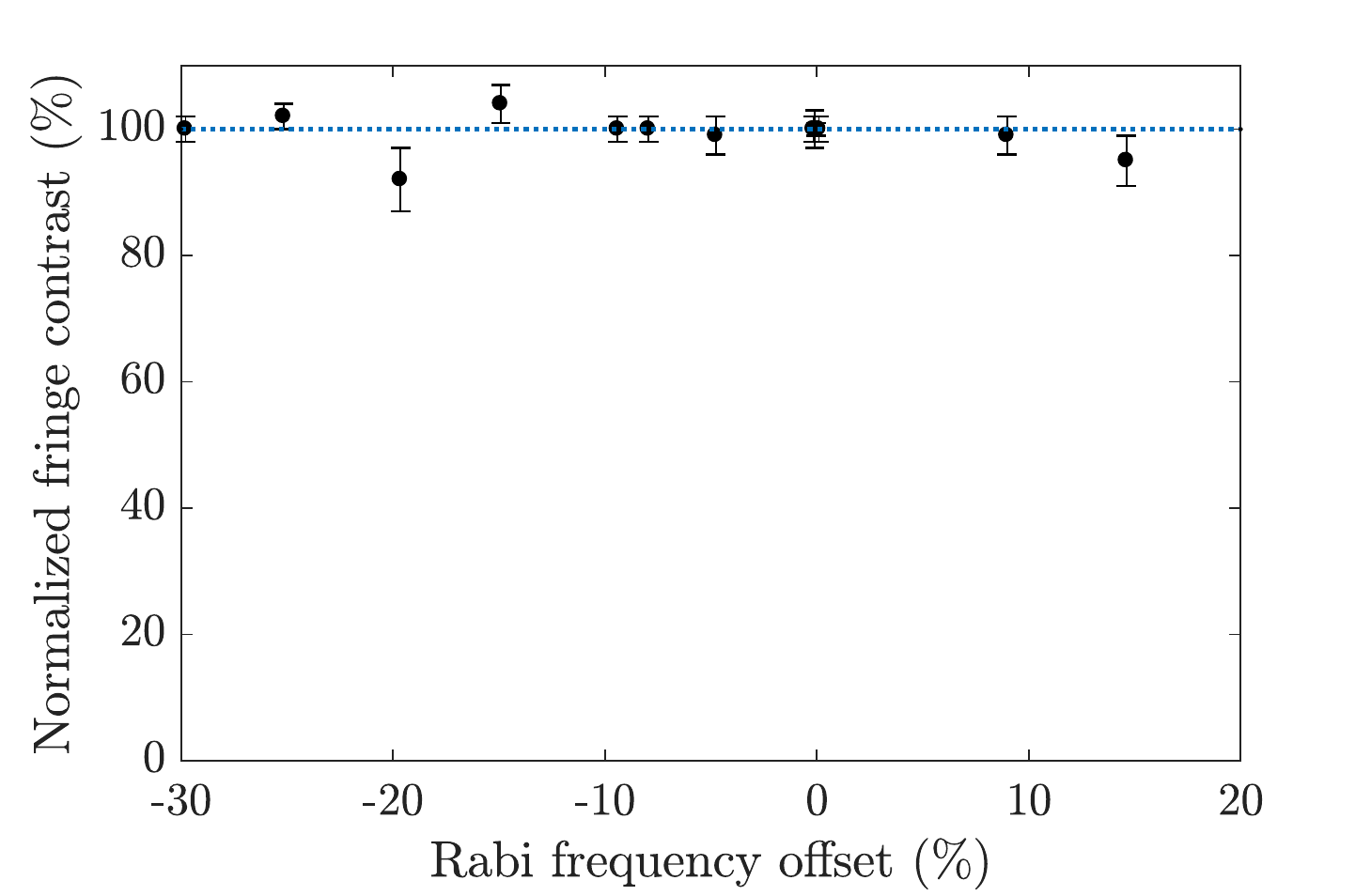}%
	\caption{}
    \end{subfigure}
\hfill
    \begin{subfigure}[b]{0.48\columnwidth}
    \includegraphics[width = \columnwidth]{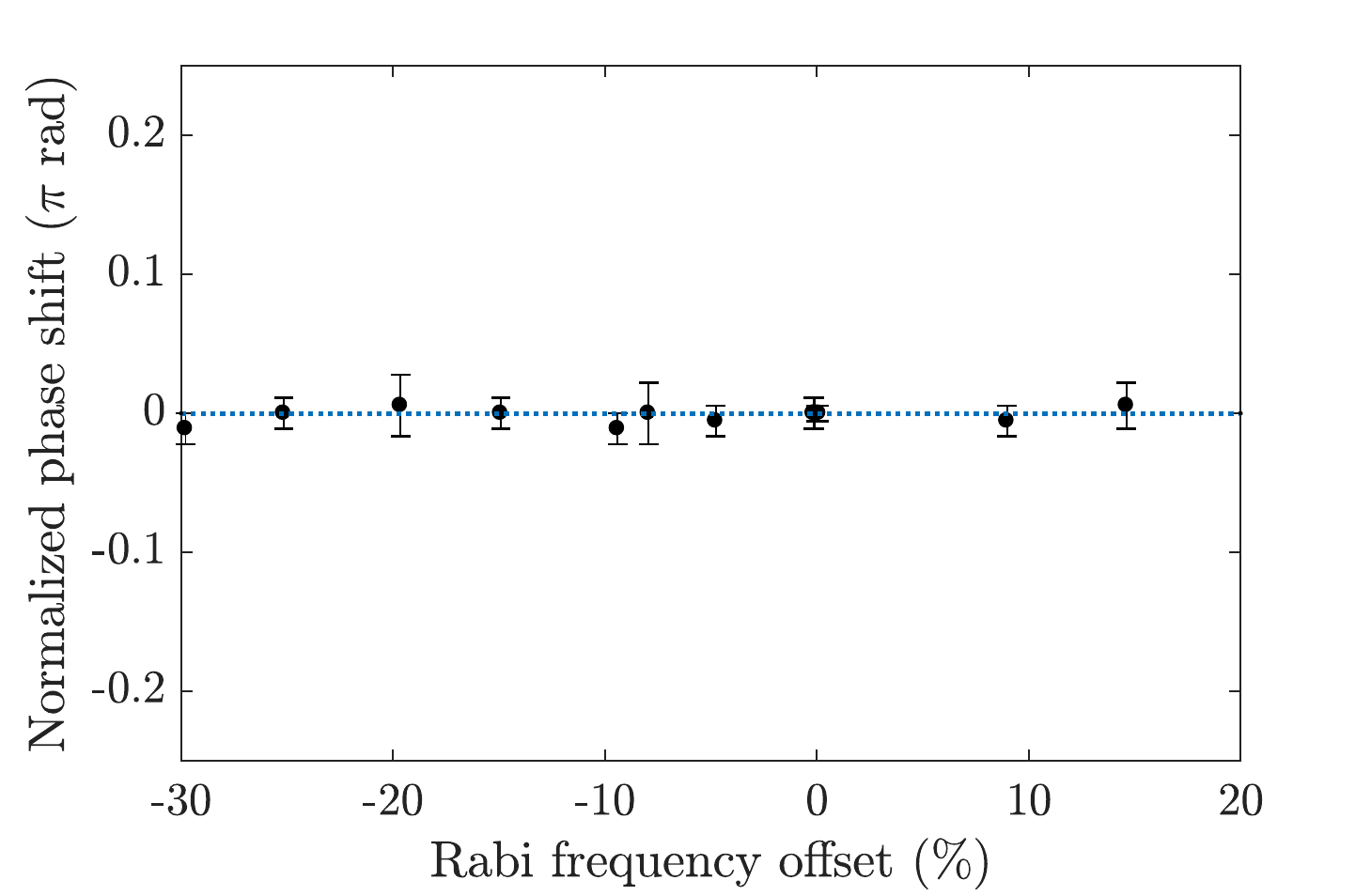}%
    \caption{}
    \end{subfigure}
\caption{\label{fig:Timing}Effect of errors in sequence timings on the resulting Ramsey fringe contrast (a) and phase shift (b) of the AXY gate, when an AXY-16 sequence is applied to generate a $\frac{\pi}{4}$ phase gate. The ions' Rabi frequencies are shifted by changing the RF amplitudes, but pulse times are re-calibrated accordingly. This way there are no pulse errors, but the time in between the pulses differs from the pre-calculated timings by the amount of the shift in Rabi frequency. The dashed line indicates a reference setting with no such errors, to which all results are normalized.}
\end{figure}

The second type of error is induced, when pulse area errors are 
minimized by re-calibration of the Rabi frequency. Pulse times are adjusted accordingly, to match the desired $\pi$ rotations. As a consequence, $\pi$ pulses are applied correctly, but deviations in pulse timings of the AXY sequence are introduced. While this could be counteracted by a recalculation of the sequence, it is not possible on experimental time scales, due to the required calculation times. 
In particular, the spacings in between pulses differ from the calculations, while their position within the sequence, defined by the parameters $\tau_a$ and $\tau_b$, are still correct.
As can be seen in Fig.~\ref{fig:Timing} there is no significant effect of such timing errors on fringe contrast or phase, even for very large errors up to 30\%.
From these results it can be seen, that matching the exact pulse times to the simulations is much less critical than matching pulse times to the applied RF power.

\section{\label{sec:Summary}Summary and outlook}

In conclusion, we have demonstrated an experimental implementation of a high-fidelity two-qubit tunable conditional phase gate in a RF-driven trapped-ion setup. The gate is generated by a specifically tuned dynamical decoupling sequence on the qubits' carrier transitions.
We also use this phase gate to implement an entangling operation shown here by generation of a Bell state.

Furthermore, the effect of motional excitation of the COM mode, errors in trap frequency as well as pulse and timing errors have been investigated. The gate is particularly robust to the ions' vibrational excitation, as up to 10~phonons on the axial COM mode do not spoil gate performance in phase nor fringe contrast. Axial trap frequency and pulse errors can have an effect on the resulting phase, while errors in sequence timings do not show any deterioration of the gate performance.
For the axial trap frequency, a gate infidelity $<10^{-4}$ is expected from the precision obtainable by tickling measurements and low pass filtering to the $10^{-5}$ regime.
Pulse errors only start to show an effect at levels $>2\%$, while state-of-the-art RF setups are capable of amplitude stabilities $<0.1\%$.
The gate implementation presented in this work is resource efficient as only a single RF field is used per ion. Even a single-tone RF field can be sufficient by redesigning pulse sequences. This is beneficial for scalability, as it avoids limitations such as amplitude resolution and storage capacity of signal generators and amplifier power.
In future setups - a planar ion trap with magnetic gradient of 120~T/m and improved coherence time has recently been put into operation - we expect to make use of both motional modes of the crystal and a corresponding speed-up of the gate time. In combination with the improved coherence times, considerably faster gates with gate times on the order of $100~\mu$s and stable fidelities $>99\%$ are expected from the respective simulations. Since the coherence time is now much longer than the gate time, the number of DD pulses can be reduced further to a minimum of a 20-pulse AXY-4 sequence in the future.

\ack
We acknowledge financial support from the EU Horizon 2020 Project 820314~(microQC) and from the German Federal Ministry of Education and Research under grant number 13N15521. 
P.B. and P.H. thank O. G\"uhne for helpful discussions. J.C. acknowledges the Ram\'on y Cajal program  (RYC2018-025197-I) as well as financial support from Spanish Government via EUR2020-112117.

% Specify following sections are appendices. Use \appendix* if there
% only one appendix.
%\appendix
%\section{}

% Create the reference section using BibTeX:
%\newpage
\section*{References}
\bibliographystyle{iopart-num}
\bibliography{AXY}

\end{document}